\documentclass[traditabstract]{aa} 

\usepackage{aalongtable}
\usepackage{longtable}
\usepackage{times,mathptm}
\usepackage{supertabular}

\usepackage{graphicx}
\usepackage{txfonts}
\newcommand{\ltsima}{$\; \buildrel$<$ \over \sim \;$}
\newcommand{\simlt}{\lower.5ex\hbox{\ltsima}}
\newcommand{\gtsima}{$\; \buildrel > \over \sim \;$}
\newcommand{\simgt}{\lower.5ex\hbox{\gtsima}}

\newcommand{\cgs}{ ${\rm erg~cm}^{-2}~{\rm s}^{-1}$} 
\newcommand{\lum}{{\rm erg~s$^{-1}$}}

\def\lesssim{\mathrel{\hbox{\rlap{\hbox{\lower4pt\hbox{$\sim$}}}\hbox{$<$}}}}
\def\gtrsim{\mathrel{\hbox{\rlap{\hbox{\lower4pt\hbox{$\sim$}}}\hbox{$>$}}}}

\def\arcmin{\hbox{$^\prime$}}
\def\arcsec{\hbox{$^{\prime\prime}$}}
\def\ks{K$_{\rm S}$}
\def\micron{\hbox{$\mu$m}}
\def\mjy{\hbox{$\mu$Jy}}

\def\ks{$K_{\rm S}$}

\def\ab1450{$AB_{1450(1+z)}$}

\def\xray{\hbox{X-ray}}

\def\mgii{\hbox{Mg\ {\sc ii}}}

\def\edd_ratio{$\log\ L_{\rm bol}/L_{\rm Edd}$}
\newcommand\phn{\phantom{0}}%
\def\asca{{\it ASCA\/}}
\def\chandra{{\it Chandra\/}}

\def\heao1{{\it HEAO-1\/}}

\def\spitzer{{\it Spitzer\/}}
\def\subaru{{\it Subaru\/}}

\def\xmm{{XMM-{\it Newton\/}}}

\def\aj{AJ}
\def\araa{ARA\&A}
\def\apj{ApJ}
\def\apjl{ApJ}
\def\apjs{ApJS}
\def\ao{Appl.~Opt.}

\def\aap{A\&A}

\def\mnras{MNRAS}

\def\procspie{Proc.~SPIE}

\begin{document}

\title{On the nature of red galaxies: the Chandra perspective}

\author{M.A. Campisi,\inst{1}
C. Vignali,\inst{2,3}
M. Brusa,\inst{4}
E. Daddi,\inst{5}
A. Comastri,\inst{3}
L. Pozzetti,\inst{3}
D.M. Alexander,\inst{6}
A. Renzini,\inst{7}
N. Arimoto\inst{8,9}
and X. Kong\inst{10}
}

\authorrunning{M.A. Campisi et al.}
\titlerunning{\chandra\ observations of EROs}

\offprints{M.A. Campisi \\email:{\tt campisi@mpa-garching.mpg.de}}

\date{Received ... ; accepted ...}

\institute{ 
Max-Planck-Institut f\"ur Astrophysik (MPA), Karl-Schwarzschild-Str. 1, 
Garching b. Muenchen, 85748, Germany
\and 
Dipartimento di Astronomia, Universit\`a degli Studi di Bologna, Via Ranzani 1, Bologna, Italy
\and
INAF--Osservatorio Astronomico di Bologna, Via Ranzani, 1, 40127 Bologna, 
Italy
\and
Max-Planck-Institut f\"ur Extraterrestrische Physik (MPE), Giessenbachstr. 1, 
85748 Garching, Germany
\and
Laboratoire AIM, CEA/DSM - CNRS - Universit\'e Paris Diderot, DAPNIA/SAP, 
Orme des Merisiers, 91191 Gif-sur-Yvette, France
\and
Department of Physics, Durham University, South Road, Durham, DH1 3LE, UK
\and
INAF -- Osservatorio Astronomico di Padova, Vicolo dell'Osservatorio 5, I-35122 Padova, Italy
\and
National Astronomical Observatory of Japan, Osawa 2-21-1, Mitaka, Tokyo 181-8588, Japan
\and
The Graduate University for Advanced Studies, Osawa 2-21-1, Mitaka, Tokyo 181-8588, Japan
\and
Center for Astrophysics, University of Science and Technology of China, Hefei 230026, China
}

\abstract{
We present the \xray\ properties of the extremely red objects (ERO) population 
observed by \chandra\ with three partially overlapping 
pointings (up to $\approx$~90~ks) over an area of $\approx$~500~arcmin$^{2}$, 
down to a 0.5--8~keV flux limit of $\approx10^{-15}$~\cgs. 
We selected EROs using a multi-band photometric catalog down to a \ks-band magnitude of $\approx$~19.3 (Vega system); 
14 EROs were detected in X-rays, corresponding to $\approx$~9\% of the overall \xray\ source 
population (149 \xray\ sources) and to $\approx$~5\% of the ERO population (288). 
The \xray\ emission of all \xray\ detected EROs is consistent with that of an active galactic nucleus (AGN) 
($\gtrsim3.5\times10^{42}$~\lum\ at photometric redshifts $z>1$), 
in agreement with previous \xray\ observations, with an indication 
of increasing absorption between the three \xray\ brightest EROs and the 11 \xray\ faintest EROs. 
We take advantage of the good spatial resolution and limited background provided 
by \chandra\ to place constraints on the population of the \xray\ undetected EROs by a 
stacking analysis. 
Their stacked emission, whose statistical significance is 5.7$\sigma$ in the observed 0.5--8~keV band, 
provides an upper limit to the average intrinsic absorption
at $z$=1 of 2.5$\times10^{22}$~cm$^{-2}$ and corresponds to a rest-frame 0.5--8~keV 
luminosity of 8.9$\times10^{41}$~\lum\ . We estimate that any accretion-related 
\xray\ emission to the stacked signal is likely ``diluted'' by emission due to hot gas 
in normal galaxies and star-formation activity in dust-enshrouded galaxies at high redshift.

\keywords{galaxies: active --- galaxies: nuclei --- quasars: general --- X-rays: galaxies} }

\maketitle

\section{Introduction}
\label{introduction}
Major efforts have been made to understand 
how the history of cosmic star formation is linked to the assembly 
of the stellar mass density (e.g., Fontana et al. 2003; see also Renzini 2006 for a review). 
These have involved challenging observational projects, mainly at 
optical and near-infrared (NIR) wavelengths, which have aimed to complete a census of the 
high-redshift galaxy population, to characterize their physical properties and to determine how galaxies form and evolve through cosmic time (e.g., Daddi et al. 2004; Caputi et al. 2005). 
Many of these studies have focused on  red galaxy populations at \hbox{$z\approx$~1--3}, which are commonly considered to be 
the progenitors of the local early-type population at high redshift. 

An efficient criterion for selecting high-redshift galaxies is related to their $R-K$ color. 
In particular, sources with ($R-K$)$\geqslant$5 or ($I-K$)$\geqslant$4 
(where magnitudes are in the Vega system), the so-called extremely red objects 
(EROs; Elston et al. 1988), 
are characterized as being optically faint but relatively bright 
in the NIR bands. Their color is consistent with that of old, passively 
evolving galaxies, observed at $z\gtrsim1$, and with that expected from dust-enshrouded 
high-redshift star-forming galaxies (e.g., Simpson et al. 2006). The two classes populate almost 
equally the ERO sample, as pointed out over the past decade by deep VLT 
spectroscopy from the K20 survey (Cimatti et al. 2002, 2003), morphological studies 
(e.g., McCarthy 2004), color selection (e.g., Pozzetti \& Mannucci 2000; Mannucci et al. 2002; 
Wilson et al. 2007; Imai et al. 2008; Dunne et al. 2009), 
radio observations (e.g., Smail et al. 2002) and spectral energy distribution fitting 
(e.g., Fang et al. 2009).
Since the first studies of large samples of EROs, it has been clear that these sources are highly clustered 
(e.g., Daddi et al. 2000; Roche et al. 2002; Georgakakis et al. 2005; Brown et al. 2005) and 
represent one of the major components of the stellar mass 
build-up at redshifts $\approx$~1--2. 

Until some years ago, the presence of an AGN in a fraction of the 
ERO population was a matter of debate, with only a few cases of detections via the presence 
of strong emission lines in the NIR/optical spectra (Pierre et al. 2001; Willott et al. 2001; 
Brusa et al. 2003) or by deep \xray\ observations (e.g., Vignali et al. 2001). 
The advent of sensitive \xray\ instruments onboard \chandra\  and \xmm\ has opened a 
new era in the investigation of the nature of EROs; in particular, in the \chandra\  
Deep Field-North (CDF-N), the fraction of EROs with \xray\ detection was estimated to be $\approx$~30\% 
(e.g., Hornschemeier et al. 2001; Alexander et al. 2002; Vignali et al. 2002). While at the 
faintest fluxes, their \xray\ emission appears consistent with non-AGN emission (star formation 
or normal elliptical galaxy emission; e.g. Brusa et al. 2002), at the brightest \xray\ fluxes, the radiation is caused by 
non-thermal processes (i.e., emission from the AGN), and in $\simgt50$\% of the AGN-related 
cases is obscured by a large amount of gas 
(with column densities of up to a few $\times10^{23}$~cm$^{-2}$). 
These results have been subsequently confirmed and expanded by many authors 
(e.g., Mainieri et al. 2002, 2005; Roche et al. 2003; Stevens et al. 2003; Mignoli et al. 2004; Brusa et al. 2005, 
hereafter B05; Severgnini et al. 2005, 2006) 
using the large amount of data in both ultra-deep and moderate-depth \xray\ surveys. 

A significant fraction of these AGN-EROs are characterized by high \xray-to-optical flux 
ratios (X/O), which has been shown to be an indicator of obscuration 
(e.g., Fiore et al. 2003, 2008; see also Fiore et al. 2009). As a consequence, the red color selection, which allows us to 
select high-redshift sources, coupled with the high X/O criterion, which is often linked to obscured 
accretion, provides a good method for identifying high-redshift, typically high-luminosity 
obscured AGN, the long-sought after Type~2 quasars (e.g., Vignali et al. 2006 and 
references therein). However, the overall number of these sources is limited by the small area 
surveyed at ultra-faint \xray\ flux limits. 

Motivated by these considerations, B05 used pointed \xmm\ 
observations in the so-called ``Daddi field'' (Daddi et al. 2000) 
to investigate the nature of \xray\ emitting EROs. 
Nine of the 257 EROs ($3.5\%$) were detected in X-rays (B05); 
X-ray emitting EROs constitute $15\%$ (9/60) of the 2--10 keV selected sample.
Given the relatively bright \xray\ fluxes reached by these observations 
(2--10 keV flux limit of 4$\times10^{-15}$~\cgs), all of the \xray\ detected EROs are probably AGN, because of 
the \xray\ luminosities derived assuming $z=1$ ($\simgt10^{43}$~\lum) and 
the possible obscuration in 7/9 of the EROs detected at hard energies.

To study the \xray\ emitting ERO population to deeper \xray\ fluxes than  
allowed by the \xmm\ observations mentioned above, the ``Daddi field'' was observed by \chandra\ 
in 2004. One of the goals consisted of detecting the faint ERO population 
and defining the average properties of the \xray\ 
individually undetected EROs by a stacking analysis, for which the \chandra\ good spatial 
resolution and low background are properly suited.\\

The paper is organized as follows: the optical/NIR catalog is presented in Sect.2, while the 
\chandra\ data, along with the \xray\ catalog, are reported in  Sect.3 and in the Appendix. 
The properties of the \xray\ detected EROs and the characterization of the 
\xray\ undetected EROs via stacking analysis are shown in  Sect.4, while  Sect.5 provides a discussion 
in the context of past \xray\ results. A summary of our results is reported in  Sect.6.

\section{The ``Daddi field'': near-infrared and optical data}
\label{daddi}
The ``Daddi field'', centered on $\alpha$=14$^{h}$49$^{m}$29$^{s}$ and $\delta$=09\degr00\arcmin00\arcsec\ 
(J2000), was observed in the \ks\ band with the {\it ESO} NTT 3.5m telescope at La Silla in 1999, using the SOFI 
camera. 
The survey in the \ks\ band covers $\approx$~700~arcmin$^{2}$ and is 85\% complete for point-like 
sources to \ks$\leq18.8$ over the entire area and to \ks$\leq19.3$ 
over an area of $\approx$~450~arcmin$^{2}$. 
The $R$-band data, down to a 5$\sigma$ limiting magnitude of $\approx$~25.4, were taken in 1998 with the 4.2m 
William Herschel Telescope in La Palma. 

Aperture photometry was computed using the {\sc SExtractor} software (Bertin \& Arnouts 1996); the output 
catalog of \ks\ sources was matched to $R$-band counterparts (or upper limits) and
contains sources with data of a signal-to-noise ratio $>$3 (in at least the \ks\ band) in a 2\arcsec\ circular 
extraction aperture (see Daddi et al. (2000) for the details about the observing runs, data analysis, and 
matched source catalog procedure). The final sample of EROs ($R-$\ks$\geq5$) comprises 
 350 sources (231 EROs with \ks$\leq18.8$, and 119 with 18.8$<$\ks$\leq$19.3).

Further optical observations in the $B$, $I$, and $z$ bands of this field were carried out 
using the \subaru\ telescope in 2003, matching the $R$ and \ks-band coverage. 
The corresponding magnitude limits in these three filters, as reported in Kong et al. (2006), 
are $\approx$~27.1, $\approx$~25.6, and $\approx$~25.0, respectively (Vega system). 
\spitzer\ IRAC and MIPS observations of this region were obtained from the 
\spitzer\ archive and reduced using {\sc moped} and customized scripts for flat fielding. 
The limiting AB magnitudes (Oke 1974) in the IRAC bands are $\approx$~24.0, 23.2, 21.4, and 20.1 
at 3.6~\micron, 4.5~\micron, 5.8~\micron, and 8.0~\micron, 
respectively, while the 24~\micron\ catalog is complete to a flux density of $\approx$~100~\mjy. 
The multicolor photometry in the $BRIzK$ and in the first three IRAC bands 
was used to estimate the 
photometric redshifts of the $K$-band selected galaxies. 
The {\sc hyperz} code (Bolzonella et al. 2000) was used 
with the four local templates from Coleman et al. (1980) augmented by a blue young 
100~Myr starburst synthetic template and a 
Calzetti et al. (2000) reddening extinction law was assumed. Comparison with a limited 
spectroscopic sample of 28 galaxies with $0<z_{spec}<2.2$ gives a median 
$\Delta z/(1+z)$ of 0.017 and a semi-interquartile range of 0.072. 
The photometric-redshift distribution for the EROs in the area covered by \chandra\ (see section \ref{ero_xprop})  
is consistent overall with that obtained in past works on spectroscopic observations of EROs 
(e.g., Cimatti et al. 2002). 
Unfortunately, the redshift information about the selected EROs is limited: only one 
ERO is detected by \chandra\ (see section \ref{xeros}), which has an $R$-band magnitude of $\approx$~23.3, 
and an optical spectrum tentatively identified as that of a broad-line AGN at $z=1.12$ 
on the basis of the \mgii\ 2798\AA\  emission line. Although this spectral line is the only line reliably detected
in the VIMOS spectrum, we note that the photometric-redshift 
solution for this source is consistent with this value ($z\approx$~1.21).

\section{Chandra data}
\label{chandra_data}
The \chandra\ observations of the ``Daddi field'' consist of a mosaic of three partially 
overlapping pointings of $\approx$~30~ks each. These observations cover a total area of 
$\approx$~500~arcmin$^{2}$; while $\approx$~45\% of this area has a nominal exposure of 
$\approx$~60~ks, $\approx$~110~arcmin$^{2}$ are close to the maximum $\approx$~90~ks exposure. 
This observational strategy was chosen as being a good compromise between the 
surveyed area (similar to the inner 11\arcmin\ of the \xmm\ observations used in B05) 
and the \xray\ flux limit. 
\begin{figure*}
\centerline{\includegraphics[width=0.9\textwidth]{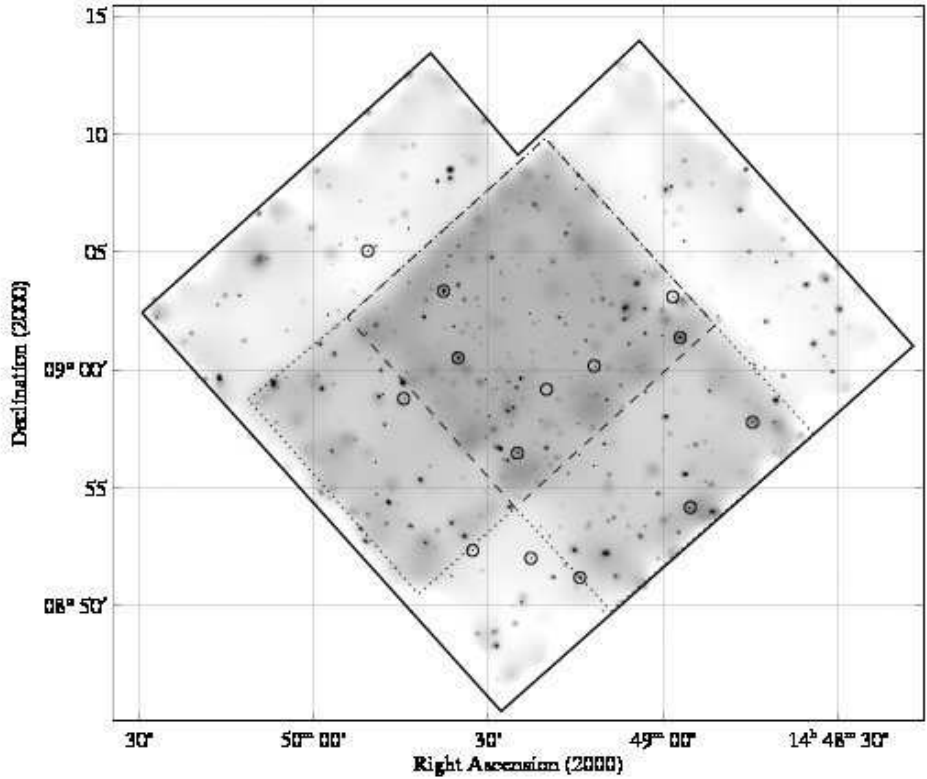}}
\caption{
\chandra\ full-band (0.5--8~keV) adaptively smoothed image of the ``Daddi field'' 
(using the algorithm of Ebeling et al. 2006), comprising 
three partially overlapping ACIS-I pointings. The open circles indicate the 
\xray\ detected EROs in these \chandra\ observations. The darker central region, bounded by dashed lines, is that 
with the deepest exposure ($\approx$~90~ks), while the intermediate-depth region ($\approx$~60~ks) lies within the dotted lines.}
\label{field}
\end{figure*}

\subsection{Data reduction}
\label{data_red}
These three observations (OBS-IDs 5032, 5033, and 5034) were performed  in June 2004 by the 
Advanced CCD Imaging Spectrometer (ACIS; Garmire et al. 2003) with the I0 CCD at the aimpoint and 
all ACIS-I CCDs in use. 
Faint mode was used for the event telemetry, and \asca\ grade 0, 2, 3, 4, and 6 events were 
used in the analysis, which was carried out using {\sc CIAO} version 3.4.2 with the latest 
relevant calibration products (CALDB 3.3.0). The event files were filtered to remove the 
limited number of flares (a few hundred seconds) and cleaned of hot pixels. 

For each observation, we produced instrument maps, exposure maps, and images  in the 
soft band \hbox{(SB=0.5--2~keV)}, hard band \hbox{(HB=2--8~keV)}, and full band \hbox{(FB=0.5--8~keV)}, 
following the recipes of the \chandra\ threads\footnote{See http://cxc.harvard.edu/ciao/threads.}. 
The exposure maps were computed using, as a spectral template, a power law with photon index $\Gamma$=1.8 
and then combined using the $merge\_all$ script and the OBS-ID=5034 observation as a reference imaging, 
in a similar way to analysis of the science images. 
The final \chandra\  mosaic in the 0.5--8~keV band is shown in Fig.~\ref{field}.
The \xray\ centroids were checked for systematic errors with respect to the optical 
positions of bright sources in the field, similarly to B05, and found that no relevant 
correction should be applied to the \xray\ data. 

Source detection was carried out with {\sc wavdetect} (Freeman et al. 2002), 
using wavelet transforms (with wavelet scale sizes of 1, 1.4, 2, 2.8, 4, 5.7, 8.0, and 11.3 
pixels) and a false-positive probability threshold of 10$^{-6}$ on the 1-arcsec binned mosaic 
image in the soft, hard, and full band, after removing the DETNAM keyword from the headers;  
this corresponds to a ``blind'' detection where counts are extracted without taking into 
account the Point Spread Function (PSF) at the source positions 
(which would represent the merging of multiple PSFs in 
the mosaic images). 
This detection procedure generated a source list in the three separate bands, 
and the resulting catalogs were 
processed by the {\sc ACIS Extract} ({\sc ae})\footnote{The {\sc ACIS Extract} software package 
and User's Guide are available online at the web site {http://www.astro.psu.edu/xray/docs/TARA/ae\_users\_guide.html.}}
software package (v. 1.31; Broos et al. 2008) 
to extract PSF-corrected source and background counts, fluxes, and \xray\ spectra. The advantage of this 
strategy consists of using, on a source-by-source basis, the effective PSF (normalized by the 
source counts) at the source position, retaining also information about the shape of the PSF. 
A comparison of the results of ``standard'' aperture photometry and {\sc ae} photometry in the 
CDF-N (using \chandra\ exposures) for a much larger source sample indicated good agreement, especially for 
sources above 40 full-band counts (F. Bauer, private communication).

\subsection{X-ray source catalog}
\label{src_catalog}
We detected 133, 92, and 154 sources in the SB, HB, and FB, respectively. 
We visually checked these sources and removed 23 sources that were clearly spurious 
(mostly PSF wings) or too close 
to chip gaps or the instrument field-of-view to derive accurate photometry. 
By matching the \xray\ sources in the three bands using a variable radius (2~arcsec in the 
innermost regions and 4 in the outermost regions) and visual inspection, we then found 138 \xray\ 
sources in the FB ($\approx$80\% of which were also detected in the SB and HB) and an additional 
3 and 8 sources with detections only in the SB and HB, respectively, 
for a total number of 149 \xray\ sources. None of these sources appears extended (as expected, given the 
detection algorithm and the scales used), although extended \xray\ emission is present in these \chandra\ 
observations as well as in the \xmm\ pointings (Finoguenov et al., in preparation). 
At first sight, the number of sources detected in only the HB may appear large compared to other fields 
(e.g., one source in the CDF-N, Alexander et al. 2003; three sources in C-COSMOS, Elvis et al. 2009), 
but we note that the detection process and the association of sources detected in different bands  
differ throughout the literature. 
This number might be a simple fluctuation; however, we note that it is not inconsistent 
with expectations from the \xray\ background synthesis models (XRB; Gilli et al. 2007) at 
the \xray\ fluxes probed by our observations. 
The sources detected only in the HB are real after visual inspection, are extremely 
faint in X-rays, and were not detected by \xmm\ because they were below the limiting flux (see B05). 

In the region in common between our mosaic and the \xmm\ observations (those actually used by B05, i.e., 
the inner 11\arcmin\ of the EPIC field-of-view), we have 138 \xray\ sources from \chandra\ and 
96 sources from \xmm\ (in the 0.5--10~keV band). 
Only 78 of the \xmm\ sources were also detected by \chandra; 
for the 18 sources detected only by \xmm, approximately two-thirds are in \chandra\ 
low-exposure regions (of below 30~ks), while the remaining sources are close to the chip gaps. 
The large number of sources detected only by \chandra\ in the region in common can be explained 
by the higher sensitivity of ACIS. 

\begin{figure}
\centerline{\includegraphics[width=88mm,angle=-90]{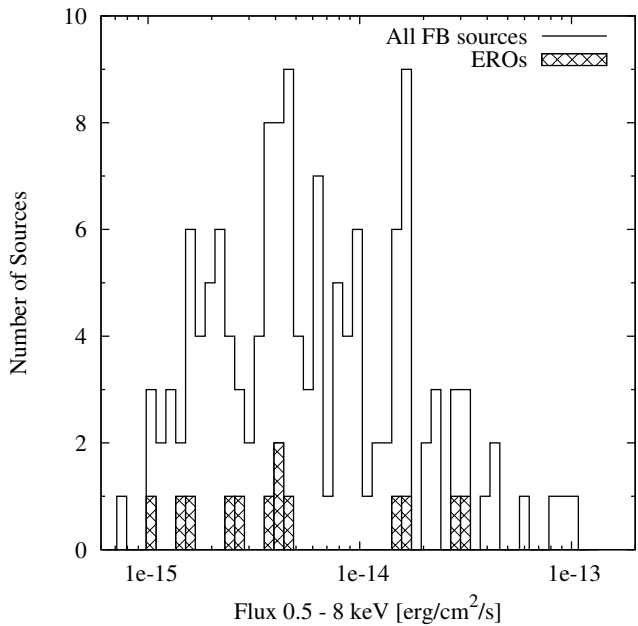}}
\caption{Full-band flux distribution for the \chandra\ sources detected in the 0.5--8~keV band, 
including the 13 EROs detected in the FB (dashed histogram).}
\label{histo_flux}
\end{figure}

For all of the \xray\ sources detected by \chandra\ and reported in the Appendix, 
we assume, as count rate-to-\xray\ flux conversion, $\Gamma=1.8$ for the FB sources, 
typical of unabsorbed AGN (e.g., Page et al. 2003; Piconcelli et al. 2005), 
while $\Gamma=2.0$ and $\Gamma=1.4$ are adopted for the sources detected only in 
the SB and HB, respectively. 
For the FB sample, we report the \hbox{0.5--8~keV} flux distribution in Fig.~\ref{histo_flux}. 
The derived FB flux limit is $\approx10^{-15}$~\cgs. 

In addition to fluxes, we can obtain approximate \xray\ spectral information by calculating the hardness 
ratio, defined as $$HR=\frac{H-S}{H+S}$$ where H and S are the count rates in the hard 
(2--8 keV) and soft (0.5--2 keV) bands, respectively. 
The results are reported in section \ref{ero_xprop} and in the Appendix. 

Using the \ks-band catalog (see section 2), we found 56 \xray\ sources with NIR counterparts 
within a matching radius of 3\arcsec. 
For the majority of the \xray\ sources, the optical counterpart is within a distance of 1\arcsec, because of the high
accuracy of the positions provided by \chandra. 

\begin{table*}
\centering
\caption{Properties of the \xray\ detected EROs}
\begin{tabular}{ccccccccccc}
\hline\hline\\ 
ID   &    XID   &   XMM ID   &     RA	   &     DEC	  &   $R$   &  $K$   &  $R-$\ks\ &  HR  &  F$_{\rm 0.5-8~keV}$ & $z_{\rm phot}$      \\ [0.05cm]
     &          &            &   (deg)     &    (deg)     &  (mag)  &  (mag) &  (mag)    &      &     (\cgs)  & \\ [0.1cm] 
\hline\\ [0.05cm]
4012 &  12 &     273    & 222.185975   & 8.962944    &  24.63  & 18.61  &  6.02	& $-0.26\pm0.50$  & 4.00E-15 & 1.52	\\  [0.05cm]         
3690 &  20 &     344    & 222.230500   & 8.902914    &  23.36  & 17.81  &  5.55	& $0.35\pm0.41$	  & 4.19E-15 & 1.11	\\  [0.05cm] 	    
3629 &  23 &     195    & 222.237642   & 9.022706    &  24.11  & 18.33  &  5.78	& $0.10\pm0.15$	  & 1.53E-14 & 1.50	\\  [0.05cm]  
3137 &  48 &            & 222.299196   & 9.002874    &  24.94  & 18.81  &  6.13	& $<-0.09$        & 1.00E-15 & 1.55	\\  [0.05cm] 	     
3009 &  51 &     239    & 222.309454   & 8.853264    &  23.96  & 18.24  &  5.72	& $-0.37\pm0.22$  & 1.73E-14 & 1.08	\\  [0.05cm]  
2760 &  64 &     237    & 222.333292   & 8.986256    &  24.93  & 19.11  &  5.82	& $<-0.15$        & 1.35E-15 & 1.62	\\  [0.05cm]          
2663 &  69 &            & 222.344513   & 8.866972    &  25.15  & 19.13  &  6.02	&                 & 2.58E-15 &          \\  [0.05cm] 	     
2591 &  73 &     293    & 222.354121   & 8.941257    &  24.26  & 19.14  &  5.12	& $0.25\pm0.30$	  & 3.64E-15 & 1.69	\\  [0.05cm] 	     
2267 &  89 &     310    & 222.386204   & 8.872362    &  24.94  & 19.07  &  5.87	& $-0.18\pm0.49$  & 4.69E-15 & 1.81	\\  [0.05cm]  
2160 &  93 &     209    & 222.396533   & 9.008494    &  23.27  & 18.09  &  5.18	& $-0.39\pm0.09$  & 2.83E-14 & 1.12$^{a}$\\ [0.05cm]  
2073 & 103 &     148    & 222.406892   & 9.055835    &  25.24  & 18.77  &  6.47	& $-0.35\pm0.09$  & 3.00E-14 & 2.53	\\  [0.05cm]  
1777 & 111 &     250    & 222.435442   & 8.979769    &  23.50  & 18.48  &  5.02	& $>$0.72         & 2.46E-15 & 1.65	\\  [0.05cm] 
1557 & 121 &            & 222.461171   & 9.084203    &  24.06  & 18.61  &  5.45	&                 & 1.66E-15 & 1.13 	\\  
\cline{1-11} \\ 
\multicolumn{11}{c}{ERO detected only in the hard band} \\ 
&&&&&&&&&F$_{\rm 2-10~keV}$ \\ [0.1cm]
3590 &      142 &            & 222.243017   & 9.051691    &  23.29  & 18.09  &  5.20    &                 & 3.76E-15 & 1.07   \\   [0.05cm]
\hline
\hline
\end{tabular}
\vskip 0.2cm
\parbox{6.3in}
{\small\baselineskip 9pt
\footnotesize
Notes --- ID and XID indicate the optical/NIR and \xray\ identification number, respectively; 
XMM ID provides the association to the B05 catalog based on \xmm\ observations. 
Coordinates are referred to the position of the associated NIR counterpart. 
Sources detected in the FB only have no HR entry. 
$^{a}$ Tentative spectroscopic redshift based on the \mgii\ 2798\AA\  emission line (see section \ref{daddi} for details). 
The photometric redshift for this source is $z_{\rm phot}$=1.21. Only one source (ID=2663) has no photometric redshift, 
because of the lack of IRAC coverage.} 
\label{tabella_eros}
\end{table*}

\section{Identification of the X-ray emitting EROs and their properties}
\label{xeros}
One purpose of this work is to establish the fraction of \xray\ emitting EROs in 
a large and clearly defined sample of relatively bright EROs and their overall properties, 
taking advantage of the moderately deep flux limit achieved by the \chandra\ exposures. 
Among the 288 EROs in the three-pointing \chandra\ mosaic, 
we found 14 EROs with \xray\ emission 
(with net counts ranging from $\approx$~5 to $\approx$~240) 
using a variable matching radius of 2\arcsec--4\arcsec\ to 
account for the PSF broadening at large off-axis angles. Visual inspection of the 
matched sources using $R$- and \ks-band images provided further support to the associations, 
most of which are at NIR-to-\xray\ distances below 1\arcsec. 
Only two sources have displacements of $\approx$~4\arcsec\, because of their large off-axis 
positions and faint \xray\ emission.  

Eight \xray\ detected EROs are associated with star-forming $BzK$ (sBzK), while one source (\#2663) is 
a passively evolving galaxy (pBzK), according to the photometric technique used by Daddi et al. (2004) to 
select star-forming and passive galaxies at $z\approx1.4-2.5$. 
We note that all of the eight EROs classified as sBzK have photometric redshifts above 1.4 
in Table~\ref{tabella_eros}, while the only pBzK has no photometric-redshift solution. 
We note that in the \chandra\ mosaic there are 122 sources classified as 
sBzK (51 of which also match the ERO definition) compared to 27 pBzK, (which are, unsurprisingly, all EROs. For 
comparison, see Greve et al. 2009). 
Among the sBzK sources, 28 ($\approx$~23\%) have \xray\ emission 
(in particular, 8 of the 51 sBzK galaxies are also classified as EROs, i.e. $\approx$~16\%
have \xray\ emission), while only one pBzK (ERO $\#2663$; see above) has \xray\ emission (i.e., 3.7\%). 
Given the flux limit reached by the \chandra\ observations, it is likely that 
most of these \xray\ emitting sBzK sources host an AGN. 
The contamination of the BzK diagram by bright AGN is a known issue 
(e.g., Daddi et al. 2004, 2007; Reddy et al. 2005; Kong et al. 2006); 
for these objects, the nuclear emission can increase the $K$-band flux significantly, 
thus placing the sources in the sBzK range. 
We have also checked whether the colors of the only \xray\ detected pBzK could be consistent, given the photometric 
uncertainties, with those of the other sBzK galaxies, but this is not the case.

Of the 14 EROs detected by \chandra\ (13 in the FB and one in the HB only; see Table~\ref{tabella_eros} 
and dashed histogram in Fig.~\ref{histo_flux}), 
all but four were also detected in the \xmm\ 
observations presented by B05 (one of these only in the soft band); the four undetected 
EROs were just below the flux limit of the \xmm\ observations. 
We note, however, that four EROs detected by both \chandra\ and \xmm\ were not 
classified as such according to the catalog used by B05 because of the slightly different optical and 
near-IR photometry. 

The EROs detected by \chandra\ represent $\approx5\%$ of the ERO sample 
(for comparison, \xmm\ detected $\approx3.5\%$) and $\approx9\%$ of the \xray\ source population 
(compared to $\approx10\%$ in the \xmm\ observations). 
Despite there being apparently similar fractions of \xray\ detected EROs between \chandra\ and \xmm, we note that the original 
optical/NIR catalog used for ERO selection in B05 and in this work differ. 
Besides that, \chandra\ is able to detect much fainter sources than \xmm\ 
but only within limited area they surveyed; 
the strength of \chandra\ is maximized for the definition of the \xray\ faint ERO properties, 
mostly due to the lower background.

\subsection{X-ray properties of the EROs}
\label{ero_xprop}
\begin{figure}
\centerline{\includegraphics[width=88mm,angle=-90]{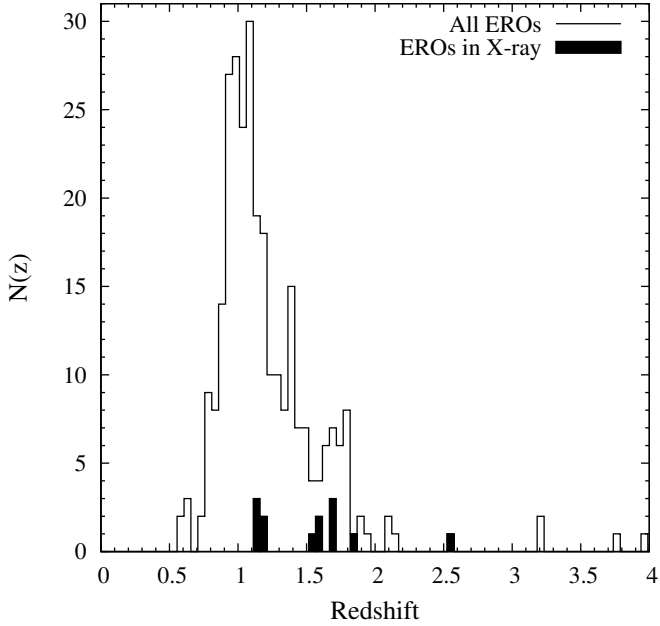}}
\caption{
Photometric-redshift distribution of the EROs in the area covered by the \chandra\ observations 
presented in this paper. The shaded histogram indicates the \xray-detected EROs.}
\label{photoz_EROs}
\end{figure}

All of the EROs associated with \xray\ sources are reported in Table~\ref{tabella_eros}, while their 
photometric-redshift distribution is shown in Fig.~\ref{photoz_EROs}. 

Using a Kolmogorov-Smirnov (KS) test, we can compare the 
photometric-redshift distribution of \xray\ detected EROs with the 
overall population in our \chandra\ mosaic (Fig.~\ref{photoz_EROs}); 
the resulting probability that the two samples are drawn from the same parent population is 
relatively low ($6.87\times10^{-3}$). 
This result is also suggested by the average redshift of the two samples: 
while EROs with \xray\ emission have $\langle z_{\rm ph} \rangle = 1.49$ 
(median $z$=1.52), the remaining EROs 
have a lower average redshift ($\langle z_{\rm ph} \rangle = 1.28$; median $z$=1.11). 
The higher redshift range of \xray\ emitting EROs might be interpreted as evidence that more luminous sources 
(hence, probably more massive galaxies, as a preliminary study of the masses of these galaxies seems to suggest) 
host an AGN more often, although the possibility that photometric 
redshifts are biased toward higher values for luminous sources is a viable explanation, as 
tentatively suggested by ERO $\#2160$. 
Clearly, the limited number of EROs with \xray\ emission in the current sample 
(and also in the literature) prevents us from drawing firm conclusions 
about this issue and demands further investigation of larger samples.

For each source, Table~\ref{tabella_eros} provides the optical/NIR and \chandra\ identification numbers (ID and XID), 
the ID in the \xmm\ catalog presented by B05, the NIR position, 
the $R$ and \ks-band magnitudes, the $R-$\ks\ color, the hardness ratio, the FB (HB) flux, 
and the photometric redshift. 

\begin{figure}
\centerline{\includegraphics[width=88mm,angle=-90]{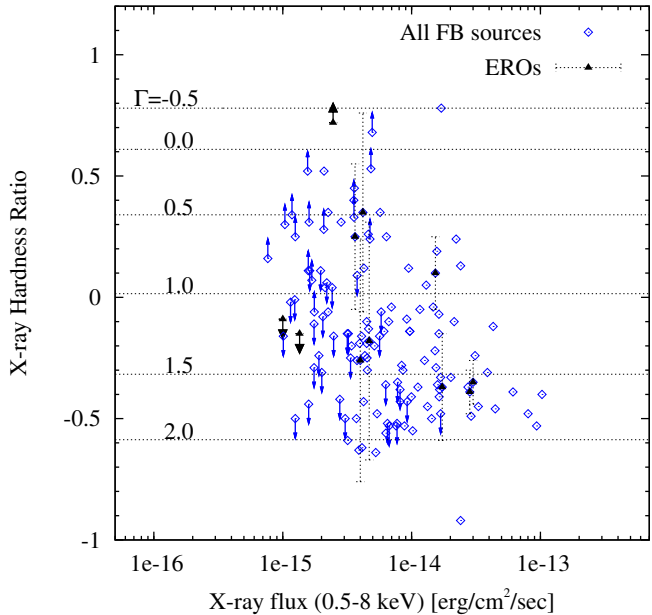}}
\caption{Hardness ratio versus 0.5--8~keV flux for all the FB sources for which a 
HR measurement was possible (open diamonds; see Appendix) and for the 
\xray\ detected EROs (filled triangles) described in Table~\ref{tabella_eros}. 
To avoid crowding, errors in the HR are reported only for EROs. 
On the left side, the photon indices corresponding to several values of HR are shown.}
\label{hr}
\end{figure}

The average (median) HR value for the EROs, after excluding the upper and lower limits 
from the computation,\footnote{Average values in the case of censored data can be calculated 
using {\sc asurv} (LaValley et al. 1992), which cannot be used when both upper and lower 
limits are present.} 
is $-0.11\pm{0.01}$ ($-0.22$), while for all other sources detected in the three bands 
it is $-0.21\pm{0.03}$ ($-0.26$). In Fig.~\ref{hr}, we report the hardness ratio versus FB flux for all the sources in the field 
(open diamonds) and for the EROs (filled triangles), along with the photon indices (horizontal dotted lines in the figure) 
corresponding to given HR values (based on the assumption of a simple power-law model with $\Gamma=1.8$). 
Below a FB flux of $\approx10^{-14}$~\cgs, we note that EROs seem to have higher HR values, 
suggesting flatter photon indices, hence possible obscuration, as the average \xray\ spectrum (section 4.2.1) appears to indicate. 

In the following, we divide the sample of \xray\ detected EROs in two subsamples, 
the former comprising the three \xray\ brightest sources ($>$100 net counts) and the latter including the remaining 
11 EROs with faint \xray\ emission. We also provide \xray\ constraints on the population 
of the individually \xray\ undetected EROs by means of a stacking analysis. 

In all of the following analyses, source counts were extracted 
using the {\sc ae} package described in section \ref{data_red}. 
The redistribution matrix files (RMFs, which include information about the detector gain and 
energy resolution) and the ancillary response files (ARFs, which include information about the 
effective area of the instrument, filter transmission and any additional energy-dependent 
efficiencies) from each observation were summed and weighted properly.

\subsection{The X-ray ``bright'' subsample}
\label{bright_sample}
We were able to extract a moderate-quality \xray\ spectrum for only three EROs, namely 
$\#2073$, $\#2160$, and $\#3629$ (see Fig.~\ref{three_Xray_spectra} in the Appendix). 

Spectral counts were accumulated in energy bins with at least 10--20 counts per bin over the 
energy range \hbox{0.5--8~keV}, although limited signal was typically present above $\approx$~4~keV. 
Spectral analysis was carried out using {\sc xspec} (version 12.4.0; Arnaud 1996) and 
the $\chi^2$ statistic. Galactic absorption (N$_{\rm H}$=2.0$\times10^{20}$~cm$^{-2}$; 
Dickey \& Lockman 1990) was included in all the spectral fittings. Errors are reported 
at the 90\% confidence level for one parameter of interest (Avni 1976).  

While the first two spectra are characterized by photon indices typical of unobscured AGN 
($\Gamma$=1.72$^{+0.28}_{-0.26}$ and $\Gamma$=1.62$^{+0.25}_{-0.23}$, respectively), 
with marginal evidence for absorption in the source $\#2160$ (rest-frame column density 
of 1.42$^{+1.92}_{-1.10}\times10^{22}$~cm$^{-2}$; in this case, the photon index would become 
$\Gamma$ $\approx$~2.0), the third source is harder, with $\Gamma$=0.82$^{+0.34}_{-0.33}$. 
For this ERO, the fit 
improves by $\Delta\chi^{2}\approx9$ for one additional parameter when the absorption is 
taken into account (N$_{\rm H}$=1.63$^{+1.05}_{-0.88}\times10^{23}$~cm$^{-2}$ at $z_{\rm phot}=2.53$; 
$\Gamma=2.19^{+0.41}_{-0.49}$). 
Assuming their photometric redshifts, these sources would have de-absorbed, rest-frame 2--10~keV 
luminosities of 11.0, 1.3, and 5.0$\times10^{44}$~\lum, respectively. 
If, on the other hand, a lower redshift ($z=1$) is assumed 
(which can be considered as a lower limit, given their $R-$\ks\ colors), 
these sources would still be in the quasar regime (i.e., above $10^{44}$~\lum\ in the 
rest-frame \hbox{2--10~keV} band).

\subsubsection{The X-ray ``faint'' subsample}
\label{weak_sample}
Given the paucity of counts for all of the remaining \xray\ detected EROs, 
we decided to investigate their average spectral properties. 
For each source, \xray\ spectra extracted from the different \chandra\ pointings 
were combined using the {\sc ftool mathpha}; 
similarly, response matrices were combined using the {\sc addrmf} and {\sc addarf} tasks. 
The total number of net counts for these 11 sources is $\approx$~170. 

A stacked spectrum was produced by deredshifting the individual spectra using the photometric 
redshifts (see Table~\ref{tabella_eros}), available for all but one of these EROs. 
Each spectrum was binned into 17 spectral channels, which correspond to those with a 
0.5~keV resolution in the rest-frame \hbox{1.7--10.2~keV} band, and corrected for the detector response curve. 
These spectra were summed together after matching the energy scale according 
to the redshift, and the data of low signal-to-noise ratio were further rebinned. 
The source with no redshift information is assumed to have a redshift $z$=1.53, the median value of the other 10 sources. 

A single power-law model provides a relatively good fit; 
the resulting photon index is flat ($\Gamma$=1.10$^{+0.42}_{-0.49}$). 
The inclusion of absorption produces a fit of comparable quality and the following 
spectral parameters: $\Gamma=1.35^{+1.63}_{-0.66}$ and an upper limit to the 
rest-frame column density of 6.4$\times10^{22}$~cm$^{-2}$ (see Fig.~\ref{ERO11_restframe_spe} in the Appendix). 
Fixing the photon index to $\Gamma$=1.8 provides a column density of 2.4$^{+2.6}_{-1.6}\times10^{22}$~cm$^{-2}$, 
which is consistent with the lack of reliable signal below 2~keV.
Assuming the spectral modeling described above, the average de-absorbed rest-frame 
2--10~keV luminosity would be $\approx4\times10^{43}$~\lum, which is clearly 
associated with AGN emission.

We note that similar spectral results, within the errors, are obtained by fitting 
the unbinned data with the Cash statistic (Cash 1979). 

\subsection{Average properties of the X-ray undetected EROs using a stacking analysis}
\label{undetected_eros}
To constrain the average \xray\ properties of the remaining 274 individually undetected EROs, 
we applied a stacking technique (e.g., Brandt et al 2001; Nandra et al. 2002; Brusa et al. 2002; 
Alexander et al. 2005). 
This technique consists of adding together the faint (if any) contribution 
from each source at the corresponding optical/NIR position 
 to estimate the average properties of the remaining ERO population, 
which probably comprises faint and obscured AGN, normal galaxies, and dust-enshrouded 
starbursts (e.g., Alexander et al. 2002). Unlike most of the \xray\ 
stacking analysis results published, we preferred to adopt a variable extraction 
radius to account, on a source-by-source basis, for the variable PSF at different off-axis 
positions in each of the three pointings, using the {\sc ae} tools. As an additional check, we also 
adopted count extraction from fixed-radius (5\arcsec) regions to verify, to first order, 
our stacking results. 

During the stacking process, all of the known (i.e., already detected) 
\xray\ sources were masked carefully (i.e., on the basis of the number of counts and 
PSF size) to avoid spurious contamination of the stacked signal by the PSF wings. 
The total exposure of the stacked EROs corresponds to $\approx$~11.9~Ms. 
Among the three bandpasses chosen for the stacking analysis, the highest signal is 
in the 0.5--2~keV band ($\approx$~6.2$\sigma$) and provides $\approx$~120 net counts; 
a $\approx$~5.7$\sigma$ signal is obtained over the entire 0.5--8~keV band ($\approx$~180 net counts). 
In the HB, a signal of $\approx$~2.5$\sigma$ is observed. 
Although deeper \xray\ observations would be useful to place stronger constrains on the 
undetected ERO population, we note that probably only deep fields 
(like the CDF-N and the \chandra\ Deep Field-South, CDF-S) 
can provide a truly significant step in this direction. 
Within the statistics of this number of source counts, we derived an average photon index 
$\Gamma=1.4^{+0.8}_{-0.6}$ and an upper limit to the column density of 2.5$\times10^{22}$~cm$^{-2}$ (at $z$=1). 
If $\Gamma$=1.8 is assumed, the \hbox{0.5--8~keV} flux ($\approx1.7\times10^{-16}$~\cgs) 
translates into a rest-frame ($z$=1) luminosity of 8.9$\times10^{41}$~\lum\ 
(1.1$\times10^{42}$~\lum\ at the median redshift $z$=1.11), 
suggestive of a likely combination of both 
accretion processes (low-luminosity and/or obscured AGN) and thermal processes for the 
majority of the ERO population. 
Similar values are obtained if the average photometric redshift for these EROs ($z$=1.28) 
is assumed. 
The results derived using fixed-radius aperture photometry are 
similar, once the statistical uncertainties are taken into account. 
However, caution is needed when results from stacking analysis of individually undetected sources are presented. 
First of all, the broad redshift range of the ERO population strongly limits the ``definition'' of the 
low-energy photo-electric cut-off (hence of the obscuration) in the stacked spectrum. 
Secondly, given the possibly broad, intrinsic column density distribution of EROs, it is plausible that 
the resulting absorption is shifted towards lower values than the true median of the sample because 
most detected counts presumably come from sources with low absorption; we note, however, 
that none of the stacked sources seems to dominate the resulting \xray\ signal.

To support our results, we also generated one-hundred catalogs of 274 fake sources, 
each randomly placed across the \chandra\ mosaic 
(once the real \xray\ sources had been masked properly), and produced stacked results 
using fixed-radius (5\arcsec) aperture photometry to assess the relevance of our stacking results 
on individually undetected EROs. 
We found that all mock catalogs but one have a number of net counts 
(source counts after background removal) below $\approx$~20, the only exception having $\approx$~100 net 
counts (i.e., lower than for our stacked EROs) in the full band. Therefore, simulations provide further 
support to the stacking analysis of the \xray\ undetected EROs. 

Since \xray-detected EROs appear to have a photometric-redshift distribution that peaks at higher values 
than the rest of the ERO population (section \ref{ero_xprop}), we divided the sample into a high-redshift and a low-redshift subsamples, 
where the dividing threshold was chosen to be $z$=1.4. Once the EROs without photometric redshift information 
had been excluded, the high- and low-redshift samples comprised 47 and 184 sources, respectively. 
For the $z>1.4$ sample, a stacked signal was detected in the FB and SB (at the 4.5$\sigma$ and 
3.7$\sigma$ significance level, respectively), while the HB signal is significant at the 2.8$\sigma$ level. 
On the basis of the FB count rate and the assumption of a power-law spectrum with $\Gamma=1.8$, we obtained a flux of 3.5$\times10^{-16}$~\cgs, 
which translates, at the source median redshift of $z$=1.72, into a rest-frame \hbox{0.5--8~keV} luminosity 
of 6.9$\times10^{42}$~\lum, consistent with AGN emission. 
We note that the stacked signal of EROs at $z>1.4$, taken at face 
value, provides a HR$=0.00\pm0.30$ which, for the hypothesis of a $\Gamma=1.8$ 
power-law model and absorption intrinsic to the source, may be interpreted as being caused by 
a column density of $\approx8.5\times10^{22}$~cm$^{-2}$ at $z=1.72$.  

On the other hand, the stacked emission from the more numerous $z<1.4$ ERO population is characterized by lower significance 
(2.8$\sigma$, 3.7$\sigma$, and 0.8$\sigma$ in the FB, SB, and HB, respectively). At the median redshift of 
$z$=1.04, the flux from the stacked FB signal (9.1$\times10^{-17}$~\cgs) is converted into a rest-frame 
\hbox{0.5--8~keV} luminosity of 5.4$\times10^{41}$~\lum, which, taken at face value, is indicative of activity unrelated 
to accretion processes (although low-luminosity AGN may be present in this ERO sample). 
The absence of signal in the HB seems to support this interpretation.

\section{Discussion}
\label{discussion}
\subsection{High X/O sources among the ERO population sampled by \chandra}
\label{high_xo}
\begin{figure}
\centerline{\includegraphics[width=88mm,angle=-90]{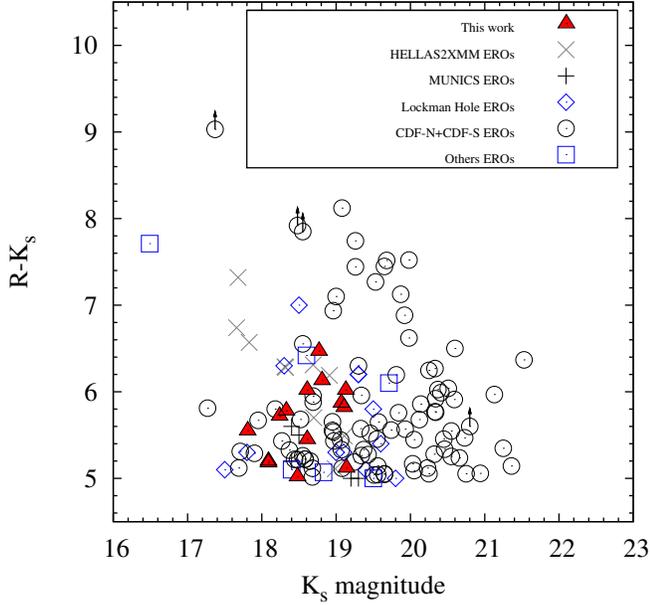}
}
\caption{$R-$\ks\ color versus \ks-band magnitude for the EROs presented in this paper and for a compilation of 
EROs from literature, similarly to B05 (HELLAS2XMM, Mignoli et al. 2004; MUNICS, Severgnini et al. 2005; 
Lockman Hole, Mainieri et al. 2002; CDF-N and CDF-S, Alexander et al. 2002, Barger et al. 2003, Szokoly et al. 2004; 
the remaining EROs, plotted as open squares, are from Crawford et al. 2002, Brusa et al. 2003, 
Willott et al. 2003, Gandhi et al. 2004, Severgnini et al. 2006; see Brusa 2004 for further details). 
In particular, all of the EROs with \xray\ detection published in this work are plotted as filled triangles.}

\label{rk_k}
\end{figure}

A comparison with data for some ERO samples from the literature shows that EROs in the ``Daddi field'' 
are not extreme, both in terms of color $R-$\ks\ (none has $R-$\ks\ $>$7) and magnitude 
(all have relatively ``bright'' \ks\ magnitudes; see Fig.~\ref{rk_k}). 
In the following, we try to provide a comprehensive picture of EROs from the \xray\ perspective, also using 
 information from available literature samples. 

\begin{figure}
\centerline{\includegraphics[width=88mm,angle=-90]{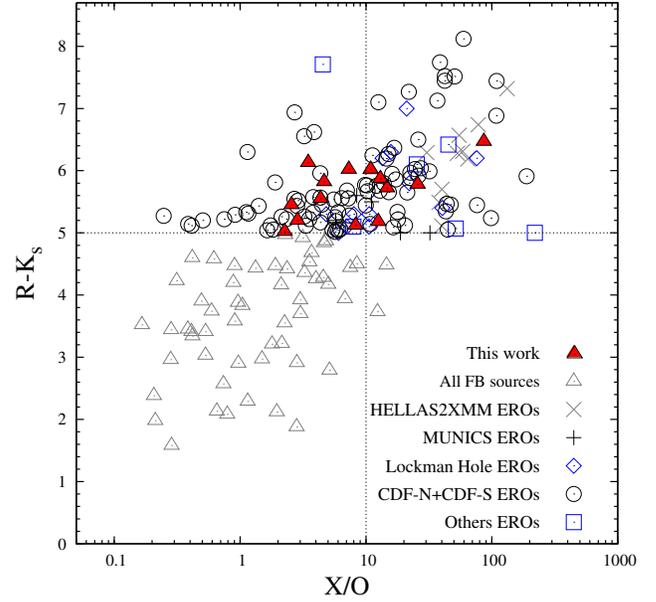}
}
\caption{
$R-$\ks\ color versus X/O ratio for all the FB sources with optical counterparts (open triangles) 
and for the 14 EROs detected by \chandra\ in the ``Daddi field'' (filled triangles). 
For comparison, EROs with \xray\ detection from literature works have also been plotted. 
The horizontal and vertical lines indicate $R-$\ks=5 and X/O=10, respectively, to guide the eye.} 
\label{x_o}
\end{figure}

A significant correlation between the properties of EROs 
and high \xray-to-optical (X/O)\footnote{The \xray-to-optical flux ratio (X/O) 
can be written (e.g., Hornschemeier et al. 2001) as: 
\begin{equation}
log \frac{f_x}{f_R}=log f_x + \frac{R}{2.5}+5.5 
\end{equation}
where $f_x$ is the flux in the X-band (0.5--8~keV in our case) and $R$ is the optical magnitude.}
flux ratio sources has been found. 
%
%
\begin{figure}
\centerline{
\includegraphics[width=88mm,angle=-90]{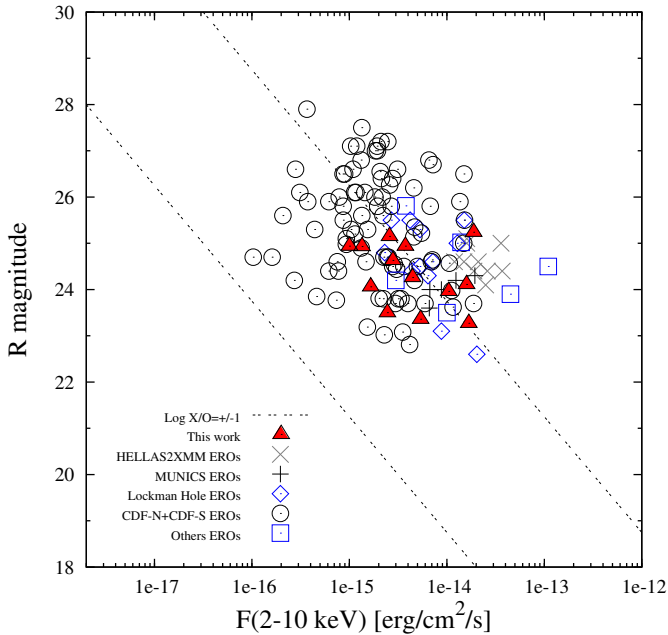}
}
\caption{$R$-band magnitude vs. 2--10~keV flux for a compilation of EROs from literature works, 
as in Fig.~\ref{rk_k}. 
Note that $\approx$~43\% of our sample (filled triangles) lie above the line of $\log(X/O)>1$ (upper dotted-dotted line).}
\label{x_o2}
\end{figure}
%
Most typical AGN have $-1<log \frac{f_x}{f_R}<1$ (i.e., lie in the region bounded by the 
two diagonal lines in Fig.~\ref{x_o2}; see Maccacaro et al. 1988), while normal galaxies and starbursts populate the 
$log \frac{f_x}{f_R}<-1$ locus (i.e., below the lower line in Fig.~\ref{x_o2}). 
The present sample of \xray\ detected EROs covers the \xray\ flux intervals of other \xmm-based 
studies (e.g., the HELLAS2XMM survey, Mignoli et al. 2004; MUNICS, Severgnini et al. 2005), 
and partially overlaps with the bright tail of the \chandra\ deep-field source distribution. 
Since high X/O objects are typically high-redshift ($z\approx1-2$), luminous ($L_{2-10~keV}>10^{44}$~\lum), 
and obscured (N$_{\rm H}>10^{22}$~cm$^{-2}$) AGN (e.g., Fiore et al. 2003), it seems 
likely that a significant fraction of EROs with high X/O are Type~2 quasars. 
In particular, as also shown in Fig.~\ref{x_o}, 6/14 EROs ($\approx$~43\%) have high X/O. Among these, we 
find the three EROs with the highest counting statistics, whose \xray\ spectral analysis (discussed in section 4.2) 
provides tentative evidence of obscuration in the source \#2160 and likely presence in the source \#3629 
(as also supported by its HR; see Table~\ref{tabella_eros}); their \xray\ luminosities are also typical of quasars. 
Overall, given the low counting statistics for most of our EROs, it is difficult to place strong constraints on 
the ERO column density distribution on the basis of the HR and to further assess the presence of obscured AGN 
among the high X/O sources.

\subsection{X-ray emission from EROs: comparison with previous work}
\label{comparison}
The \xray\ analysis carried out in this work has shown that a fraction of EROs, 
mostly at faint \xray\ fluxes, is probably obscured, although no indication of Compton-thick absorption (i.e., of column density above 
$\approx$~10$^{24}$~cm$^{-2}$; e.g., Comastri 2004) has been found in the combined spectrum of 
11 \xray\ weak EROs (see section \ref{weak_sample}). 
This result confirms previous findings for \chandra\ and \xmm\ observations of EROs (e.g., 
Alexander et al. 2002; Vignali et al. 2002; Stevens et al. 2003; B05; Severgnini et al. 2005; Del Moro et al. 2009), i.e., 
the presence of obscuring matter in a significant fraction of \xray\ detected EROs, at least 
above $\approx$~10$^{-15}$~\cgs\ (see Brusa 2004 for details). 
This is confirmed by Fig.~\ref{x_o}, where the X/O ratio is plotted versus the $R-$\ks\ color; 
in this figure, $\approx$~50\% of EROs have X/O$>$10. 
One, or possibly two, of the \xray\ brightest EROs (in the quasar regime) in the ``Daddi field" 
are likely to be obscured (with column density of up to $\approx~10^{23}$~cm$^{-2}$ for source \#3629), 
thus being Type~2 quasars; similarly, the average \xray\ properties of the population of 
\xray\ weak EROs are consistent with obscured AGN emission 
 (N$_{\rm H}\approx$~2.4$^{+2.6}_{-1.6}\times10^{22}$~cm$^{-2}$, assuming $\Gamma=1.8$).
Also the hardness-ratio analysis provides evidence of obscuration, although with even larger uncertainties: 
the average HR for all of the \xray\ detected EROs (excluding upper and lower limits) is larger than for the remaining \xray\ sources. 
Besides the uncertainties in deriving column densities from sources with limited redshift information, we 
note that high signal-to-noise \xray\ observations of Seyfert galaxies and quasars suggest that 
the ``single power-law assumption'', typically adopted in studies based on HR, may be too crude to reproduce properly 
the \xray\ emission (hence, the spectral parameters) of our AGN, although the 
likely redshift above $z$=1 strongly limits the influence of any soft excess in the observed energy band. 

The high sensitivity of the \chandra\ observations to point-like source detection, 
coupled with their high spatial resolution and low background, have also allowed us to place constraints  
on the average properties of the individually undetected EROs by a stacking analysis. 
Although at faint \xray\ flux limits, we expect to find a higher contribution to the \xray\ emission 
from passively evolved galaxies and dusty-enshrouded galaxies (e.g., Alexander et al. 2002; 
Brusa et al. 2002; Roche et al. 2006; see also Ranalli et al. 2003), 
our stacking analysis seems to identify a non-negligible contribution from obscured and/or low-luminosity 
AGN emission; 
a model fit to the stacked spectrum of \xray\ undetected EROs indeed provides an upper limit 
of \hbox{$2.5\times 10^{22}$~cm$^{-2}$} to the column density (at $z$=1) and 
an overall source luminosity of $8.9\times10^{41}$~\lum.

\section{Summary}
\label{summary}
We have presented three \chandra\ observations in the ``Daddi field'' ($\approx$~30~ks exposure each), 
down to a 0.5--8~keV flux limit of $\approx10^{-15}$~\cgs. 
Here is a summary of our main results:\\

$\bullet$ We have detected 149 \xray\ sources, 14 of which are classified as EROs (13 detected in the FB and 
one in the HB only). The fraction of \xray\ detected EROs corresponds to $\approx$~9\% of the overall \xray\ source 
population and to $\approx$~5\% of the ERO population. \\

$\bullet$ Under the reasonable assumption that EROs are located at $z\gtrsim1$, 
the \xray\ emission of all of the \xray\ detected EROs is consistent with their being AGN 
($L_{\rm X}\gtrsim3.5\times10^{42}$~\lum\ at $z>1$), at least the three \xray\ brightest EROs 
lying in the quasar luminosity range (i.e., above 10$^{44}$~\lum). \\

$\bullet$ Moderate-quality \xray\ spectral analysis was possible overall for the three \xray\ brightest EROs 
and the 11 \xray\ faintest EROs. Spectral results are consistent with 
obscuration in one, possibly two, of the \xray\ brightest EROs (of column density as high as 10$^{23}$~cm$^{-2}$), 
and in the subsample of \xray\ weak EROs 
($\approx2.4^{+2.6}_{-1.6}\times10^{22}$~cm$^{-2}$), assuming $\Gamma=1.8$. \\

$\bullet$ We also took advantage of the good spatial resolution and limited background provided 
by the \chandra\ data to place constraints on the population of the 274 \xray\ undetected EROs in the 
\chandra\ observations by means of a stacking analysis. 
Modeling the stacked spectrum of these EROs with an absorbed power law provides an upper limit 
of \hbox{$2.5\times 10^{22}$~cm$^{-2}$} to the column density (at $z$=1); 
the derived \xray\ signal corresponds to a rest-frame \hbox{0.5--8~keV} luminosity of 8.9$\times10^{41}$~\lum.
It seems likely that the underlying ERO population consists of: (i) normal galaxies; (ii) bright dusty starburst galaxies, 
whose emission is related to star-forming processes (see Ranalli et al. 2003); and (iii) accretion-powered sources,  
providing support for \xray\ stacking analyses on deeper fields.
To first order, this result is confirmed by stacking each of the $z>1.4$ and $z<1.4$ ERO populations separately;  
while the higher redshift EROs have a stacked signal of significance 4.5$\sigma$ and 3.7$\sigma$ 
in the FB and SB, respectively (2.8$\sigma$ in the HB), and a corresponding rest-frame \hbox{0.5--8~keV} 
luminosity well above $10^{42}$~\lum\ at the median $z$=1.72 (hence consistent, on average, with AGN emission), 
the lower redshift ERO population is formally detected above 3$\sigma$ only in the SB (2.8$\sigma$ in the FB) 
and its average flux corresponds to a \hbox{0.5--8~keV} luminosity of 5.4$\times10^{41}$~\lum. 
This value suggests a prevailing contribution from emission mechanisms unrelated to nuclear accretion, 
albeit some low-luminosity AGN may be present as well.
%

\begin{acknowledgements}
The authors would like to thank F.E. Bauer and P. Broos for their support to the {\sc ae} package, 
K. Iwasawa for providing the rest-frame spectrum of \xray\ faint EROs, 
G. Lanzuisi for help with \xmm\ data analysis, P. Ranalli for suggestions on \xray\ stacking analysis, 
P. Severgnini for thoughtful discussion and providing data points for Fig.~\ref{x_o}, 
G. Zamorani for useful discussion, and the referee for his/her careful reading of the manuscript 
and good suggestions. 
CV thanks D. Wilman for the great organization of the CNOC2 meeting in 2007 
and the MPA/MPE institutes for their kind hospitality over the last two years. 
CV and AC acknowledge financial support from the Italian Space Agency 
(contracts ASI--INAF I/023/05/0 and ASI I/088/06/0) and PRIN--MIUR (grant 2006-02-5203);  
ED acknowledges the support of French ANR-08-JCJC-0008, and DMA of the Royal Society. 
NA is supported by a Grant-in-Aid for Science Research from JSPS (No.19540245). 
\end{acknowledgements}


\clearpage
\onecolumn

\begin{appendix}
\section{\xray\ source catalog in the ``Daddi field'' and \xray\ spectra of EROs}
\vspace{1cm}

\footnotesize
\begin{center}
\topcaption{X-ray properties of the sources detected by \chandra\ in the ``Daddi field''}

\begin{supertabular}{cccccccccc}

\hline \hline \\ 
\multicolumn{1}{c}{XID} & 
\multicolumn{1}{c}{RA} & 
\multicolumn{1}{c}{DEC} & 
\multicolumn{1}{c}{$\theta$} & 
\multicolumn{1}{c}{Counts} &
\multicolumn{1}{c}{Expo} & 
\multicolumn{1}{c}{F(0.5--8~keV)} & 
\multicolumn{1}{c}{F(0.5--2~keV)} & 
\multicolumn{1}{c}{F(2--8~keV)} & 
\multicolumn{1}{c}{HR} \\
\multicolumn{1}{c}{} & 
\multicolumn{1}{c}{(deg)} & 
\multicolumn{1}{c}{(deg)} & 
\multicolumn{1}{c}{(\arcmin)} & 
\multicolumn{1}{c}{(0.5--8~keV)} & 
\multicolumn{1}{c}{(ks)} & 
\multicolumn{1}{c}{(\cgs)} &
\multicolumn{1}{c}{(\cgs)} &
\multicolumn{1}{c}{(\cgs)} &
\multicolumn{1}{c}{} \\  [0.1cm]
\hline\noalign{\smallskip} 
{\phn\phn}1   &    222.123840 & 9.042500  &  9.5 &  39.4  &  22.2 & 1.63E-14  &   6.00E-15    &   1.41E-14    &   $-0.07\pm0.27$  \\ [0.05cm]
{\phn\phn}2   &    222.128250 & 8.998556  &  9.3 &  22.4  &  22.2 & 9.25E-15  &   4.98E-15    & $<$5.29E-15   & $<-0.43$          \\ [0.05cm]
{\phn\phn}3   &    222.133453 & 9.063889  &  9.2 &  24.0  &  22.6 & 9.71E-15  &   3.80E-15    &   7.84E-15    &   $-0.14\pm0.36$  \\ [0.05cm]  
{\phn\phn}4   &    222.142288 & 9.018167  &  8.4 &  41.6  &  23.2 & 1.65E-14  &   7.75E-15    &   9.71E-15    &   $-0.38\pm0.24$  \\ [0.05cm]
{\phn\phn}5   &    222.147797 & 9.060167  &  8.3 &  36.6  &  22.1 & 1.52E-14  &   6.36E-15    &   1.11E-14    &   $-0.22\pm0.27$  \\ [0.05cm]
{\phn\phn}6   &    222.169083 & 8.978500  &  8.4 & 105.7  &  33.4 & 2.90E-14  &   1.48E-14    &   1.39E-14    &   $-0.49\pm0.14$  \\ [0.05cm]
{\phn\phn}7   &    222.169205 & 9.075861  &  7.4 &  12.3  &  23.2 & 4.84E-15  & $<$7.96E-16   &   7.47E-15    & $>0.53$           \\ [0.05cm]
{\phn\phn}8   &    222.169998 & 8.962806  &  9.0 &  20.9  &  37.2 & 5.14E-15  &   2.13E-15    &   3.83E-15    &   $-0.20\pm0.44$  \\ [0.05cm]
{\phn\phn}9   &    222.176331 & 9.033833  &  6.4 &  19.3  &  22.9 & 7.71E-15  &   4.83E-15    & $<$4.34E-15   & $<-0.52$          \\ [0.05cm]
{\phn}10      &    222.177078 & 9.001167  &  6.5 &   9.6  &  24.7 & 3.56E-15  & $<$7.45E-16   &   5.01E-15    & $>0.40$           \\ [0.05cm]
{\phn}11      &    222.184799 & 9.003861  &  6.0 &   8.9  &  25.2 & 3.23E-15  &   1.88E-15    & $<$3.88E-15   & $<-0.15$          \\ [0.05cm]
{\phn}12      &    222.186707 & 8.962855  &  8.4 &  17.8  &  40.7 & 4.00E-15  &   1.72E-15    &   2.80E-15    &   $-0.26\pm0.50$  \\ [0.05cm]
{\phn}13      &    222.188461 & 9.083556  &  6.6 &  15.5  &  21.9 & 6.50E-15  &   3.57E-15    & $<$3.17E-15   & $<-0.52$          \\ [0.05cm]
{\phn}14      &    222.190536 & 9.141583  &  8.9 &  31.9  &  17.4 & 1.68E-14  &   1.04E-14    & $<$9.81E-15   & $<-0.48$          \\ [0.05cm] 
{\phn}15      &    222.194290 & 9.112861  &  7.5 &  31.0  &  24.5 & 1.16E-14  &   4.12E-15    &   1.06E-14    &   $-0.05\pm0.30$  \\ [0.05cm] 
{\phn}16      &    222.210922 & 9.037639  &  4.4 &   5.6  &  26.7 & 1.93E-15  & $<$6.85E-16   & $<$1.99E-15   &                   \\ [0.05cm] 
{\phn}17      &    222.214828 & 8.899639  &  8.8 &  81.5  &  43.9 & 1.70E-14  &   1.29E-15    &   2.84E-14    &   $0.78\pm0.14$   \\ [0.05cm] 
{\phn}18      &    222.217667 & 8.985028  &  6.4 &  16.5  &  44.7 & 3.36E-15  &   1.62E-15    & $<$2.71E-15   & $<-0.25$          \\ [0.05cm] 
{\phn}19      &    222.227463 & 8.947083  &  6.6 &  19.2  &  46.6 & 3.77E-15  &   1.60E-15    &   2.72E-15    &   $-0.26\pm0.40$  \\ [0.05cm] 
{\phn}20      &    222.231247 & 8.902083  &  8.1 &  20.6  &  45.1 & 4.19E-15  &   9.26E-16    &   5.38E-15    &   $0.35\pm0.41$   \\ [0.05cm] 
{\phn}21      &    222.233994 & 8.926028  &  7.0 &  79.9  &  47.4 & 1.55E-14  &   6.76E-15    &   1.06E-14    &   $-0.29\pm0.17$  \\ [0.05cm] 
{\phn}22      &    222.235077 & 9.090778  &  4.9 &  15.6  &  23.4 & 6.10E-15  &   2.33E-15    &   5.18E-15    &   $-0.14\pm0.44$  \\ [0.05cm]
{\phn}23      &    222.237701 & 9.022778  &  6.8 & 116.0  &  69.6 & 1.53E-14  &   4.71E-15    &   1.59E-14    &   $0.10\pm0.15$   \\ [0.05cm] 
{\phn}24      &    222.240295 & 9.047250  &  5.1 &   8.5  &  44.6 & 1.75E-15  &   1.09E-15    & $<$2.47E-15   & $<-0.11$          \\ [0.05cm] 
{\phn}25      &    222.244080 & 9.129583  &  6.7 &  17.4  &  25.3 & 6.30E-15  &   3.10E-15    & $<$4.12E-15   & $<-0.36$          \\ [0.05cm]    
{\phn}26      &    222.247833 & 8.966861  &  5.2 &  86.0  &  48.2 & 1.64E-14  &   7.77E-15    &   9.45E-15    &   $-0.41\pm0.15$  \\ [0.05cm]
{\phn}27      &    222.248123 & 9.127222  &  6.5 &  43.6  &  25.5 & 1.57E-14  &   4.32E-15    &   1.80E-14    &   $0.19\pm0.24$   \\ [0.05cm]
{\phn}28      &    222.249084 & 8.893778  &  7.7 &  18.4  &  36.6 & 4.61E-15  &   1.16E-15    &   5.53E-15    &   $0.26\pm0.42$   \\ [0.05cm]
{\phn}29      &    222.250549 & 9.023861  &  6.4 &  14.2  &  66.2 & 1.97E-15  &   8.25E-16    & $<$2.85E-15   & $<$0.11           \\ [0.05cm]
{\phn}30      &    222.253876 & 8.912417  &  6.8 &  22.6  &  48.8 & 4.24E-15  &   2.06E-15    &   2.32E-15    &   $-0.43\pm0.35$  \\ [0.05cm]
{\phn}31      &    222.262787 & 8.984556  &  5.1 &  24.4  &  59.8 & 3.73E-15  &   1.89E-15    &   1.82E-15    &   $-0.50\pm0.34$  \\ [0.05cm]
{\phn}32      &    222.262878 & 9.025694  &  5.9 &  20.4  &  65.8 & 2.84E-15  &   6.65E-16    &   3.57E-15    &   $0.31\pm0.40$   \\ [0.05cm]
{\phn}33      &    222.263046 & 9.069306  &  5.5 &   9.4  &  51.3 & 1.68E-15  & $<$7.90E-16   &   2.64E-15    & $>0.07$           \\ [0.05cm]
{\phn}34      &    222.263214 & 9.043528  &  6.2 &  55.8  &  60.9 & 8.38E-15  &   3.67E-15    &   5.76E-15    &   $-0.28\pm0.22$  \\ [0.05cm]
{\phn}35      &    222.265961 & 8.882333  &  7.9 &  26.7  &  38.7 & 6.32E-15  &   3.36E-15    &   2.66E-15    &   $-0.56\pm0.32$  \\ [0.05cm]
{\phn}36      &    222.268753 & 9.060694  &  6.4 & 129.7  &  70.9 & 1.68E-14  &   7.57E-15    &   1.08E-14    &   $-0.33\pm0.13$  \\ [0.05cm]
{\phn}37      &    222.273209 & 9.097361  &  6.0 &  19.9  &  40.1 & 4.54E-15  &   2.00E-15    &   3.07E-15    &   $-0.30\pm0.39$  \\ [0.05cm]
{\phn}38      &    222.274124 & 9.018278  &  5.0 &  32.8  &  76.9 & 3.90E-15  &   2.15E-15    &   1.40E-15    &   $-0.63\pm0.28$  \\ [0.05cm]
{\phn}39      &    222.277084 & 9.043833  &  5.6 & 254.3  &  70.6 & 3.30E-14  &   1.62E-14    &   1.76E-14    &   $-0.45\pm0.09$  \\ [0.05cm]
{\phn}40      &    222.277878 & 9.035528  &  6.3 &  56.2  &  58.5 & 8.79E-15  &   4.56E-15    &   4.00E-15    &   $-0.53\pm0.20$  \\ [0.05cm]
{\phn}41      &    222.284210 & 8.854389  &  8.8 &  28.7  &  41.2 & 6.38E-15  &   1.65E-15    &   7.46E-15    &   $0.25\pm0.33$   \\ [0.05cm]
{\phn}42      &    222.285797 & 8.895667  &  7.1 &  16.7  &  34.1 & 4.49E-15  &   1.67E-15    &   3.90E-15    &   $-0.10\pm0.44$  \\ [0.05cm]
{\phn}43      &    222.288589 & 8.925778  &  5.2 &  10.5  &  47.9 & 2.01E-15  &   9.77E-16    & $<$1.47E-15   & $<-0.31$          \\ [0.05cm]
{\phn}44      &    222.290375 & 8.978639  &  4.9 &  83.4  &  76.7 & 9.96E-15  &   4.74E-15    &   5.73E-15    &   $-0.41\pm0.16$  \\ [0.05cm]
{\phn}45      &    222.291122 & 8.870472  &  8.0 & 180.3  &  38.4 & 4.30E-14  &   1.64E-14    &   3.59E-14    &   $-0.12\pm0.11$  \\ [0.05cm]
{\phn}46      &    222.292206 & 8.947667  &  4.2 & 132.4  &  50.5 & 2.40E-14  &   1.54E-14    &   1.99E-15    &   $-0.92\pm0.06$  \\ [0.05cm]
{\phn}47      &    222.293457 & 9.050139  &  6.5 &  12.5  &  54.7 & 2.08E-15  & $<$6.97E-16   &   3.58E-15    & $>0.28$           \\ [0.05cm]
{\phn}48      &    222.298370 & 9.002167  &  5.2 &   5.4  &  49.6 & 1.00E-15  &   6.98E-16    & $<$1.66E-15   & $<-0.09$          \\ [0.05cm]
{\phn}49      &    222.299667 & 8.932222  &  4.7 &  13.0  &  43.2 & 2.77E-15  &   1.38E-15    & $<$1.63E-15   & $<-0.42$          \\ [0.05cm]
{\phn}50      &    222.305252 & 9.020639  &  4.4 &  10.1  &  80.2 & 1.15E-15  &   4.76E-16    & $<$1.31E-15   & $<-0.02$          \\ [0.05cm]
{\phn}51      &    222.309830 & 8.853306  &  7.0 &  46.8  &  24.8 & 1.73E-14  &   8.05E-15    &   1.04E-14    &   $-0.37\pm0.22$  \\ [0.05cm]
{\phn}52      &    222.310379 & 8.939861  &  4.5 &  10.1  &  52.7 & 1.75E-15  &   1.10E-15    & $<$1.74E-15   & $<-0.29$          \\ [0.05cm]
{\phn}53      &    222.310455 & 8.961611  &  4.9 &  13.0  &  74.4 & 1.60E-15  & $<$4.11E-16   &   2.28E-15    & $>0.31$           \\ [0.05cm]
{\phn}54      &    222.311172 & 9.137444  &  8.6 &  20.3  &  42.8 & 4.34E-15  &   1.85E-15    &   3.10E-15    &   $-0.24\pm0.45$  \\ [0.05cm]
{\phn}55      &    222.311874 & 9.127639  &  8.3 &  11.1  &  46.3 & 2.20E-15  &   1.14E-15    & $<$3.43E-15   & $<0.06$           \\ [0.05cm]
{\phn}56      &    222.313416 & 8.872611  &  7.4 &  65.5  &  46.2 & 1.30E-14  &   4.21E-15    &   1.30E-14    &   $0.05\pm0.20$   \\ [0.05cm]
{\phn}57      &    222.313416 & 9.012250  &  4.1 &  25.7  &  73.9 & 3.19E-15  &   1.71E-15    &   1.29E-15    &   $-0.59\pm0.31$  \\ [0.05cm]
{\phn}58      &    222.317245 & 9.102750  &  7.2 &  17.5  &  66.2 & 2.42E-15  &   1.05E-15    & $<$3.11E-15   & $<0.04$           \\ [0.05cm]
{\phn}59      &    222.319458 & 9.095917  &  7.1 &  28.4  &  72.2 & 3.60E-15  &   6.78E-16    &   4.99E-15    &   $0.45\pm0.31$   \\ [0.05cm]
{\phn}60      &    222.319702 & 9.032389  &  4.5 &  34.9  &  78.2 & 4.09E-15  &   1.60E-15    &   3.35E-15    &   $-0.16\pm0.28$  \\ [0.05cm]
{\phn}61      &    222.322952 & 8.994083  &  4.1 &  10.6  &  77.8 & 1.25E-15  & $<$4.39E-16   &   2.17E-15    & $>0.25$           \\ [0.05cm]
{\phn}62      &    222.324493 & 9.036972  &  4.6 &  36.4  &  77.9 & 4.27E-15  &   1.27E-15    &   4.68E-15    &   $0.12\pm0.27$   \\ [0.05cm]
{\phn}63      &    222.328293 & 8.853472  &  6.7 &  28.6  &  23.5 & 1.12E-14  &   5.18E-15    &   6.79E-15    &   $-0.37\pm0.30$  \\ [0.05cm]
{\phn}64      &    222.333252 & 8.986194  &  4.0 &  11.5  &  77.7 & 1.35E-15  &   7.43E-16    & $<$1.57E-15   & $<-0.15$          \\ [0.05cm]
{\phn}65      &    222.333328 & 9.136556  &  8.9 &  10.5  &  54.5 & 1.76E-15  & $<$8.02E-16   &   1.95E-15    & $>-0.06$          \\ [0.05cm]
{\phn}66      &    222.333923 & 8.944556  &  4.7 &  70.2  &  70.0 & 9.18E-15  &   3.36E-15    &   8.21E-15    &   $-0.09\pm0.19$  \\ [0.05cm]
{\phn}67      &    222.337250 & 8.964194  &  4.5 &  10.4  &  60.5 & 1.58E-15  & $<$5.05E-16   &   1.84E-15    & $>0.11$           \\ [0.05cm]
{\phn}68      &    222.341919 & 9.116611  &  7.9 &  34.1  &  67.4 & 4.64E-15  &   1.79E-15    &   3.83E-15    &   $-0.13\pm0.31$  \\ [0.05cm]
{\phn}69      &    222.344254 & 8.866611  &  5.7 &   7.3  &  25.8 & 2.58E-15  & $<$1.18E-15   & $<$2.04E-15   &                   \\ [0.05cm]
{\phn}70      &    222.345673 & 9.039556  &  4.6 &   8.7  &  76.9 & 1.04E-15  & $<$3.49E-16   &   1.90E-15    & $>0.30$           \\ [0.05cm]
{\phn}71      &    222.351044 & 8.835861  &  7.5 &  18.9  &  22.2 & 7.78E-15  &   4.56E-15    & $<$6.02E-15   & $<-0.35$          \\ [0.05cm]
{\phn}72      &    222.351868 & 8.993889  &  4.0 & 371.8  &  76.3 & 4.46E-14  &   2.19E-14    &   2.38E-14    &   $-0.46\pm0.07$  \\ [0.05cm]
{\phn}73      &    222.354004 & 8.941389  &  4.7 &  29.8  &  74.8 & 3.64E-15  &   9.23E-16    &   4.46E-15    &   $0.25\pm0.30$   \\ [0.05cm]
{\phn}74      &    222.355118 & 8.987278  &  3.9 &  44.7  &  77.6 & 5.27E-15  &   2.90E-15    &   1.88E-15    &   $-0.64\pm0.20$  \\ [0.05cm]
{\phn}75      &    222.356873 & 8.973278  &  3.8 &  69.9  &  74.7 & 8.57E-15  &   3.74E-15    &   5.90E-15    &   $-0.30\pm0.18$  \\ [0.05cm]
{\phn}76      &    222.359833 & 8.903028  &  5.7 &  87.9  &  49.5 & 1.63E-14  &   6.33E-15    &   1.34E-14    &   $-0.15\pm0.17$  \\ [0.05cm]
{\phn}77      &    222.360703 & 8.991722  &  3.8 &   6.8  &  81.3 & 7.66E-16  & $<$2.25E-16   &   9.24E-16    & $>0.16$           \\ [0.05cm]
{\phn}78      &    222.360840 & 8.971000  &  3.7 &  25.8  &  77.4 & 3.05E-15  &   1.85E-15    & $<$1.77E-15   & $<-0.50$          \\ [0.05cm]
{\phn}79      &    222.360916 & 8.935917  &  5.0 &  27.4  &  73.3 & 3.42E-15  &   1.38E-15    &   2.68E-15    &   $-0.20\pm0.33$  \\ [0.05cm]
{\phn}80      &    222.363464 & 9.049944  &  5.0 &   9.6  &  74.6 & 1.18E-15  & $<$3.61E-16   &   2.15E-15    & $>0.34$           \\ [0.05cm]
{\phn}81      &    222.366379 & 9.068861  &  5.7 &  10.3  &  75.3 & 1.25E-15  &   9.34E-16    & $<$8.79E-16   & $<-0.50$          \\ [0.05cm]
{\phn}82      &    222.366455 & 8.944361  &  5.5 & 200.2  &  59.0 & 3.11E-14  &   1.30E-14    &   2.29E-14    &   $-0.24\pm0.10$  \\ [0.05cm]
{\phn}83      &    222.369370 & 8.804972  &  9.4 &  50.1  &  20.7 & 2.22E-14  &   5.79E-15    &   2.56E-14    &   $0.24\pm0.22$   \\ [0.05cm]
{\phn}84      &    222.371368 & 8.814944  &  8.8 &  14.3  &  22.6 & 5.81E-15  &   2.99E-15    & $<$7.06E-15   & $<-0.06$          \\ [0.05cm]
{\phn}85      &    222.372208 & 9.079472  &  6.3 &  35.6  &  60.4 & 5.40E-15  &   2.71E-15    &   2.70E-15    &   $-0.48\pm0.26$  \\ [0.05cm]
{\phn}86      &    222.374298 & 8.966694  &  3.8 &  10.3  &  66.4 & 1.42E-15  & $<$5.52E-16   & $<$1.59E-15   &                   \\ [0.05cm]
{\phn}87      &    222.376877 & 8.993111  &  4.0 &   8.5  &  77.0 & 1.01E-15  &   4.44E-16    & $<$9.13E-16   & $<-0.16$          \\ [0.05cm]
{\phn}88      &    222.382050 & 8.992194  &  4.0 &  13.5  &  77.5 & 1.59E-15  &   8.27E-16    & $<$9.06E-16   & $<-0.44$          \\ [0.05cm]
{\phn}89      &    222.386246 & 8.872472  &  5.7 &  13.1  &  25.6 & 4.69E-15  &   1.86E-15    &   3.75E-15    &   $-0.18\pm0.49$  \\ [0.05cm]
{\phn}90      &    222.392044 & 8.882361  &  6.9 &  75.9  &  43.6 & 1.59E-14  &   7.35E-15    &   9.86E-15    &   $-0.36\pm0.17$  \\ [0.05cm]
{\phn}91      &    222.392090 & 9.049778  &  5.0 &  17.8  &  77.8 & 2.09E-15  &   3.42E-16    &   3.08E-15    &   $0.52\pm0.39$   \\ [0.05cm]
{\phn}92      &    222.396042 & 9.087750  &  6.6 &  15.7  &  70.4 & 2.05E-15  &   1.04E-15    & $<$2.47E-15   & $<-0.08$          \\ [0.05cm]
{\phn}93      &    222.396500 & 9.008444  &  4.3 & 238.9  &  77.2 & 2.83E-14  &   1.33E-14    &   1.68E-14    &   $-0.39\pm0.09$  \\ [0.05cm]
{\phn}94      &    222.396835 & 9.060028  &  5.7 &  30.5  &  70.6 & 3.95E-15  &   1.60E-15    &   3.07E-15    &   $-0.19\pm0.30$  \\ [0.05cm]
{\phn}95      &    222.398087 & 9.012417  &  4.4 &  34.8  &  77.2 & 4.12E-15  &   2.26E-15    &   1.52E-15    &   $-0.62\pm0.25$  \\ [0.05cm]
{\phn}96      &    222.399750 & 9.030722  &  4.5 &  30.9  &  63.1 & 4.48E-15  &   1.80E-15    &   3.54E-15    &   $-0.19\pm0.30$  \\ [0.05cm]
{\phn}97      &    222.401077 & 8.891778  &  6.5 &  24.0  &  44.4 & 4.95E-15  & $<$5.53E-16   &   8.33E-15    & $>0.68$           \\ [0.05cm]
{\phn}98      &    222.401382 & 8.854250  &  7.0 &   8.7  &  22.3 & 3.57E-15  & $<$8.29E-16   &   4.68E-15    & $>0.33$           \\ [0.05cm]
{\phn}99      &    222.402206 & 9.080778  &  6.4 &  18.7  &  76.0 & 2.25E-15  &   4.95E-16    &   2.91E-15    &   $0.35\pm0.40$   \\ [0.05cm]
100           &    222.402206 & 9.141528  &  7.3 & 258.9  &  23.1 & 1.03E-13  &   4.89E-14    &   5.90E-14    &   $-0.40\pm0.09$  \\ [0.05cm]
101           &    222.402588 & 9.135417  &  6.9 &  41.1  &  25.6 & 1.47E-14  &   5.20E-15    &   1.35E-14    &   $-0.04\pm0.25$  \\ [0.05cm]
102           &    222.404160 & 9.073222  &  6.1 &  15.1  &  72.5 & 1.91E-15  &   1.10E-15    & $<$1.88E-15   & $<-0.24$          \\ [0.05cm]
103           &    222.406876 & 9.055667  &  5.2 & 236.8  &  72.3 & 3.00E-14  &   1.37E-14    &   1.89E-14    &   $-0.35\pm0.09$  \\ [0.05cm]
104           &    222.411285 & 9.133194  &  6.7 &  25.6  &  24.3 & 9.66E-15  &   3.75E-15    &   7.97E-15    &   $-0.14\pm0.33$  \\ [0.05cm]
105           &    222.419876 & 8.877861  &  7.5 &  31.6  &  43.5 & 6.65E-15  &   2.50E-15    &   5.69E-15    &   $-0.10\pm0.31$  \\ [0.05cm]
106           &    222.420868 & 9.037028  &  5.7 &  85.8  &  62.2 & 1.26E-14  &   5.42E-15    &   8.94E-15    &   $-0.26\pm0.16$  \\ [0.05cm]
107           &    222.423752 & 8.922472  &  5.4 & 130.3  &  49.7 & 2.40E-14  &   7.03E-15    &   2.65E-14    &   $0.13\pm0.13$   \\ [0.05cm]
108           &    222.429993 & 8.889306  &  7.3 &  29.0  &  35.0 & 7.58E-15  &   3.94E-15    &   3.41E-15    &   $-0.53\pm0.30$  \\ [0.05cm]
109           &    222.432129 & 9.099889  &  4.5 &  16.7  &  22.8 & 6.70E-15  &   3.49E-15    & $<$3.10E-15   & $<-0.53$          \\ [0.05cm]
110           &    222.433624 & 8.998611  &  5.0 &  10.1  &  75.0 & 1.24E-15  &   6.81E-16    & $<$1.85E-15   & $<-0.01$          \\ [0.05cm]
111           &    222.435333 & 8.979750  &  3.8 &  14.3  &  53.3 & 2.46E-15  & $<$2.29E-16   &   4.22E-15    & $>0.72$           \\ [0.05cm]
112           &    222.436493 & 8.990611  &  4.6 & 398.0  &  59.6 & 6.12E-14  &   2.88E-14    &   3.62E-14    &   $-0.39\pm0.07$  \\ [0.05cm]
113           &    222.446289 & 8.906417  &  6.7 & 189.1  &  44.7 & 3.88E-14  &   1.73E-14    &   2.56E-14    &   $-0.31\pm0.10$  \\ [0.05cm]
114           &    222.449036 & 8.961111  &  4.8 &  11.9  &  51.4 & 2.13E-15  &   6.90E-16    &   2.16E-15    &   $0.04\pm0.53$   \\ [0.05cm]
115           &    222.449921 & 8.990222  &  4.0 &   9.2  &  54.2 & 1.56E-15  & $<$3.38E-16   &   3.12E-15    & $>0.52$           \\ [0.05cm]
116           &    222.451797 & 8.999694  &  3.9 &  33.0  &  53.1 & 5.70E-15  &   1.25E-15    &   7.53E-15    &   $0.35\pm0.27$   \\ [0.05cm]
117           &    222.453247 & 8.981194  &  4.2 & 110.9  &  50.4 & 2.02E-14  &   9.02E-15    &   1.33E-14    &   $-0.33\pm0.14$  \\ [0.05cm]
118           &    222.455078 & 8.911806  &  6.8 &  12.2  &  45.2 & 2.48E-15  &   1.40E-15    & $<$2.78E-15   & $<-0.16$          \\ [0.05cm]
119           &    222.455215 & 9.031222  &  5.7 &  56.6  &  74.3 & 6.97E-15  &   2.47E-15    &   6.36E-15    &   $-0.04\pm0.22$  \\ [0.05cm]
120           &    222.455826 & 8.934750  &  5.9 &   9.0  &  50.8 & 1.62E-15  &   7.71E-16    & $<$2.71E-15   & $<0.11$           \\ [0.05cm]
121           &    222.461044 & 9.083861  &  3.8 &   4.7  &  26.0 & 1.66E-15  & $<$9.39E-16   & $<$3.13E-15   &                   \\ [0.05cm]
122           &    222.462250 & 9.025194  &  6.0 &  18.1  &  73.5 & 2.25E-15  &   8.17E-16    &   2.00E-15    &   $-0.06\pm0.48$  \\ [0.05cm]
123           &    222.462372 & 8.878500  &  8.9 &  47.6  &  20.3 & 2.14E-14  &   8.14E-15    &   1.79E-14    &   $-0.10\pm0.24$  \\ [0.05cm]
124           &    222.466049 & 8.887278  &  8.3 &  31.9  &  30.8 & 9.48E-15  &   2.87E-15    &   9.99E-15    &   $0.12\pm0.30$   \\ [0.05cm]
125           &    222.480576 & 9.088806  &  4.6 &   4.8  &   9.3 & 4.76E-15  & $<$1.97E-15   &   9.36E-15    & $>0.24$           \\ [0.05cm]
126           &    222.485123 & 9.011889  &  5.3 &  70.0  &  44.8 & 1.43E-14  &   7.28E-15    &   6.89E-15    &   $-0.50\pm0.17$  \\ [0.05cm]
127           &    222.487000 & 9.073972  &  4.2 &   7.7  &  27.4 & 2.56E-15  & $<$8.92E-16   & $<$2.57E-15   &                   \\ [0.05cm]
128           &    222.487045 & 8.943139  &  6.8 &  24.1  &  48.9 & 4.52E-15  &   1.93E-15    &   3.22E-15    &   $-0.25\pm0.36$  \\ [0.05cm]
129           &    222.489670 & 9.038917  &  3.2 &  39.8  &  27.3 & 1.33E-14  &   6.52E-15    &   7.16E-15    &   $-0.45\pm0.23$  \\ [0.05cm]
130           &    222.494003 & 8.986694  &  6.2 & 143.8  &  48.1 & 2.74E-14  &   1.28E-14    &   1.64E-14    &   $-0.37\pm0.12$  \\ [0.05cm]
131           &    222.495285 & 9.018000  &  5.9 &  53.4  &  48.1 & 1.02E-14  &   5.35E-15    &   4.41E-15    &   $-0.55\pm0.19$  \\ [0.05cm]
132           &    222.496750 & 8.995306  &  6.3 &  16.0  &  46.3 & 3.17E-15  &   1.48E-15    & $<$2.96E-15   & $<-0.15$          \\ [0.05cm]
133           &    222.528961 & 8.990222  &  7.9 & 405.6  &  46.3 & 8.02E-14  &   4.09E-14    &   3.90E-14    &   $-0.48\pm0.07$  \\ [0.05cm]
134           &    222.539291 & 9.110583  &  7.9 &  14.7  &  23.9 & 5.64E-15  &   2.52E-15    & $<$4.94E-15   & $<-0.16$          \\ [0.05cm]
135           &    222.553131 & 9.052889  &  7.1 &  18.5  &  20.8 & 8.14E-15  &   4.07E-15    & $<$4.92E-15   & $<-0.38$          \\ [0.05cm]
136           &    222.567291 & 8.994028  &  7.9 & 236.7  &  23.2 & 9.35E-14  &   4.92E-14    &   4.13E-14    &   $-0.53\pm0.08$  \\ [0.05cm]
137           &    222.587585 & 9.026056  &  8.9 &   9.2  &  22.3 & 3.78E-15  &   1.92E-15    & $<$5.99E-15   & $<0.09$           \\ [0.05cm]
138           &    222.612335 & 9.036944  & 10.4 &  18.7  &  20.9 & 8.16E-15  &   4.10E-15    &   4.18E-15    &   $-0.43\pm0.47$  \\ [0.05cm]
\hline \hline 
\end{supertabular}
\begin{longtable}{ccccccc}
\hline \hline \\ 
\multicolumn{7}{c}{\sf Soft-band only sources} \\
\hline \\ [0.1cm]
\multicolumn{1}{c}{XID} & 
\multicolumn{1}{c}{RA} & 
\multicolumn{1}{c}{DEC} & 
\multicolumn{1}{c}{$\theta$} & 
\multicolumn{1}{c}{Counts} & 
\multicolumn{1}{c}{Expo} & 
\multicolumn{1}{c}{F(0.5--2~keV)} \\ [0.05cm]
\multicolumn{1}{c}{} & 
\multicolumn{1}{c}{(deg)} & 
\multicolumn{1}{c}{(deg)} & 
\multicolumn{1}{c}{(\arcmin)} & 
\multicolumn{1}{c}{(0.5--2~keV)} & 
\multicolumn{1}{c}{(ks)} & 
\multicolumn{1}{c}{(\cgs)} \\ [0.1cm]
\hline \\ 
139           &    222.328827 & 9.075750  &  6.0 &   8.9  &  75.1 & 7.21E-16 \\ [0.05cm]
140           &    222.533173 & 9.078806  &  6.5 &   7.7  &  24.5 & 1.92E-15 \\ [0.05cm]
141           &    222.563248 & 9.052389  &  7.7 &   6.1  &  23.1 & 1.61E-15 \\ [0.05cm]
%
\hline \hline \\ 
\multicolumn{7}{c}{\sf Hard-band only sources} \\
\hline \\ [0.1cm]
\multicolumn{1}{c}{XID} & 
\multicolumn{1}{c}{RA} & 
\multicolumn{1}{c}{DEC} & 
\multicolumn{1}{c}{$\theta$} & 
\multicolumn{1}{c}{Counts} & 
\multicolumn{1}{c}{Expo} & 
\multicolumn{1}{c}{F(2--10~keV)} \\
\multicolumn{1}{c}{} & 
\multicolumn{1}{c}{(deg)} & 
\multicolumn{1}{c}{(deg)} & 
\multicolumn{1}{c}{(\arcmin)} & 
\multicolumn{1}{c}{(2--8~keV)} & 
\multicolumn{1}{c}{(ks)} & 
\multicolumn{1}{c}{(\cgs)} \\  [0.1cm]
\hline \\ 
142           &    222.242661 & 9.051778 &   4.7 &   5.6  &  38.8 & 3.76E-15 \\ [0.05cm]
143           &    222.402618 & 9.103194  &  6.3 &   6.5  &  43.0 & 3.94E-15 \\ [0.05cm]
144           &    222.416534 & 8.932056  &  4.9 &   5.8  &  49.6 & 3.07E-15 \\ [0.05cm]
145           &    222.519257 & 8.994639  &  7.4 &   8.0  &  50.6 & 4.04E-15 \\ [0.05cm]
146           &    222.526703 & 8.962861  &  8.2 &   8.5  &  47.7 & 4.59E-15 \\ [0.05cm]
147           &    222.264465 & 8.930778  &  5.7 &   4.0  &  49.5 & 2.12E-15 \\ [0.05cm]
148           &    222.371552 & 8.957608  &  4.1 &   8.2  &  75.9 & 2.84E-15 \\ [0.05cm]
149           &    222.305862 & 8.995367  &  5.1 &  10.3  &  53.9 & 5.01E-15 \\ [0.05cm]
\hline
\end{longtable}
\parbox{4.5in}
{\small
\footnotesize
XID: X-ray identification number in the \chandra\ mosaic; 
(RA, DEC): X-ray source coordinates; 
$\theta$: off-axis angle at the source position averaged over the \chandra\ observations; 
Counts: source net counts; 
Expo: exposure time at the source position (from the exposure maps; see $\S$\ref{src_catalog} for details); 
F(0.5--8~keV)/F(0.5--2~keV)/F(2--8~keV)/F(2--10~keV): flux obtained from the count-rate conversion, assuming a 
power-law model with $\Gamma$=1.8/2.0/1.4/1.4, as described in the text; 
HR: hardness ratio. The upper limits (at the 95\% confidence level) to the counts, hence the 
reported fluxes, were computed according to Kraft, Burrows, \& Nousek (1991). 
Only five sources are detected in the full band without 
being detected in the two sub-bands considered in this work (because of the 
adopted detection threshold).
}

\end{center}

\begin{figure}
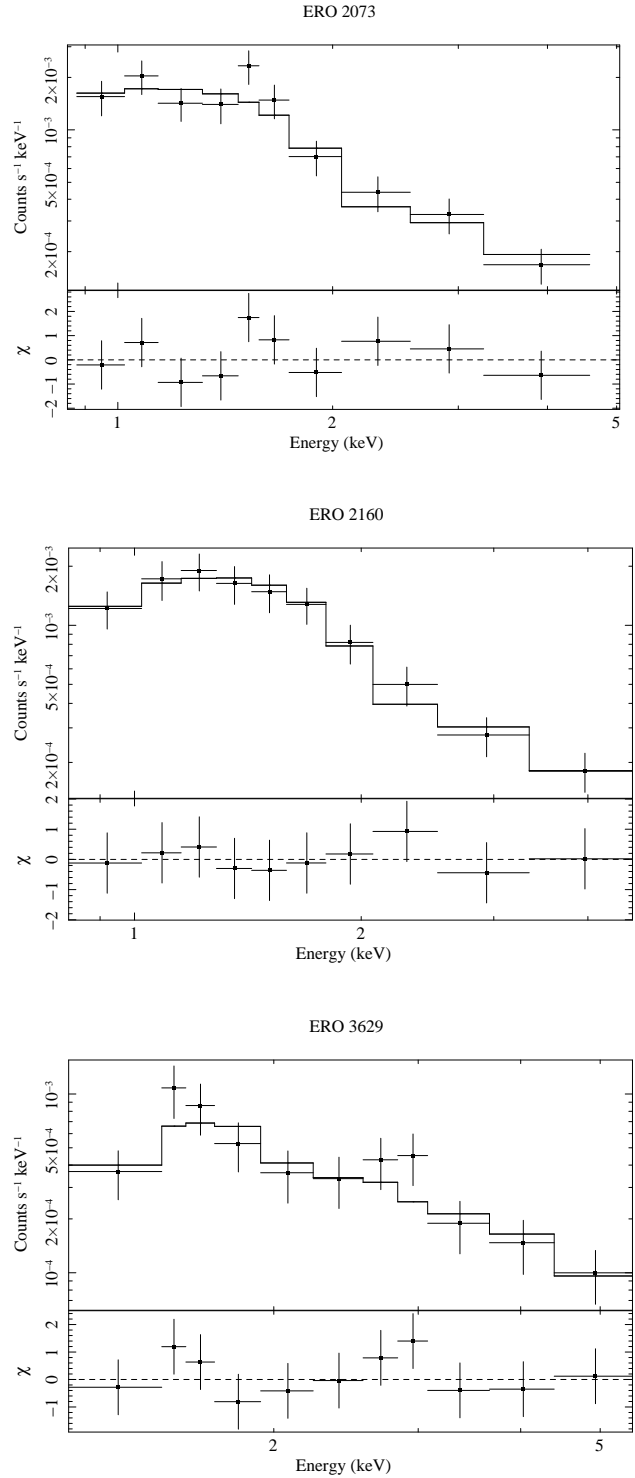

\centerline{\includegraphics[angle=-90,width=0.45\textwidth]{ero2073_spectrum.ps}}
\vglue0.6cm
\centerline{\includegraphics[angle=-90,width=0.45\textwidth]{ero2160_spectrum.ps}}
\vglue0.6cm
\centerline{\includegraphics[angle=-90,width=0.45\textwidth]{ero3629_spectrum.ps}}
\caption{
\chandra\ spectra (folded with the response of the ACIS-I instrument) 
of the three \xray\ EROs with most counts in the 0.5--8~keV band. 
While for the spectrum of ERO~\#2073 a single power law model was assumed, 
for the remaining two objects absorption at the source rest-frame (using the spectroscopic 
redshift for ERO~\#2160 and the photometric redshift for ERO~\#3629; see Table~\ref{tabella_eros}) 
was included in the spectral fitting, as described in section \ref{bright_sample}.
The three bottom panels illustrate the data/model ratio in units of $\sigma$.}
\label{three_Xray_spectra}
\end{figure}
%
\begin{figure}
\centerline{\includegraphics[angle=-90,width=0.45\textwidth]{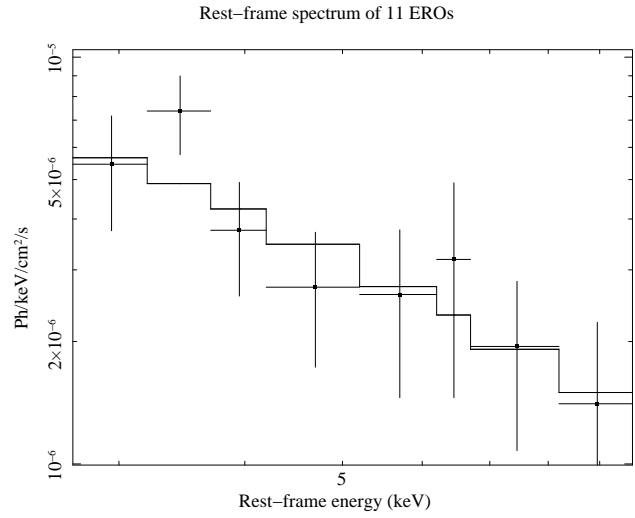}}
\caption{
Rest-frame \xray\ spectrum of the 11 EROs that are individually detected in the \chandra\ observations. 
The underlying model consists of an absorbed power law.}
\label{ERO11_restframe_spe}
\end{figure}
\end{appendix}

\end{document}